\documentclass[12pt]{iopart}

\usepackage{iopams}
\usepackage{graphicx}
\usepackage[svgnames,table]{xcolor}
\usepackage[utf8]{inputenc}
\usepackage{appendix}
\usepackage{fouriernc}
\usepackage{float}
\usepackage{hyperref}
\usepackage{subcaption}
\hypersetup{
        colorlinks=true,
        citecolor=SteelBlue,
        filecolor=LimeGreen,
        linkcolor=SlateBlue,
        urlcolor=Purple
}

\def\Msun{M_{\odot}}

\graphicspath{{figures/}}

\usepackage{cleveref}
\usepackage{lineno}
\begin{document}
\title[LIGO-India]{The Science Case for LIGO-India}

\author{M. Saleem$^{1,7}$,  Javed Rana$^{9}$, V. Gayathri$^{5,6}$,  Aditya Vijaykumar$^{2}$, Srashti Goyal$^{2}$, Surabhi Sachdev$^{9}$, Jishnu Suresh$^{11}$, S. Sudhagar$^{4}$,  Arunava Mukherjee$^{8}$, Gurudatt Gaur$^{12}$, Bangalore Sathyaprakash$^{9}$, Archana Pai $^5$, Rana X Adhikari$^{2,3}$, P. Ajith$^2$, Sukanta Bose$^{4,10}$}

\address{$^1$ Chennai Mathematical Institute, Siruseri 603103, Tamilnadu, India}
\address{$^2$ International Centre for Theoretical Sciences, Tata Institute of Fundamental Research, Bangalore 560089, India}
\address{$^3$ LIGO Laboratory, California Institute of Technology, USA}
\address{$^4$ Inter-University Centre for Astronomy and Astrophysics (IUCAA), Post Bag 4, Ganeshkhind, Pune 411 007, India}
\address{$^5$ Department of Physics, Indian Institute of Technology Bombay, Powai, Mumbai 400 076, India}
\address{$^6$ Department of Physics, University of Florida, PO Box 118440, Gainesville, FL 32611-8440, USA}
\address{$^7$ School of Physics and Astronomy, University of Minnesota, Minneapolis, MN 55455, USA}
\address{$^8$ Saha Institute of Nuclear Physics, HBNI, 1/AF Bidhannagar, Kolkata-700064, India}
\address{$^9$ Institute for Gravitation and the Cosmos, The Pennsylvania State University, University Park, PA 16802, USA}
\address{$^{10}$ Department of Physics and Astronomy, Washington State University, 1245 Webster, Pullman, WA 99164-2814, USA}
\address{$^{11}$ Institute for Cosmic Ray Research (ICRR), The University of Tokyo, Kashiwa City, Chiba 277-8582, Japan}
\address{$^{12}$ Institute of Advanced Research, Gandhinagar 382 426, Gujarat, India}
\ead{sukanta@iucaa.in}

\vspace{7pt}
\begin{indented}
\item[]\today
\end{indented}

\begin{abstract}
  The global network of gravitational-wave detectors has completed three observing runs with $\sim$50 detections of merging compact binaries. A third LIGO detector, with comparable astrophysical reach, is to be built in India (LIGO-Aundha) and expected to be operational during the latter part of this decade.
Such additions to the network increase the number of baselines and the network SNR of GW events. These enhancements help improve the sky-localization of those events.
%and the cross-correlation search for the stochastic GW backgrounds
Multiple detectors simultaneously in operation will also increase the baseline duty factor,
thereby, leading to an improvement in the detection rates and, hence, the completeness of surveys.
In this paper, we quantify the improvements due to the expansion of the LIGO Global
Network (LGN)
in the precision with which source properties will be measured.
%and quantitatively discuss how the improvements can give better astrophysical insights about the source properties and how that improves our knowledge of
We also present examples of how this expansion will give a boost to tests of
fundamental physics.

% and cosmology.
%We find that the addition of a new detector in India has substantial effects on the scientific capabilities of the LIGO Global Network.
\end{abstract}

%\maketitle

%   remove before submission

\tableofcontents

\maketitle

%=========================================================
\section{Introduction}
\label{sec:intro}

The global network of gravitational-wave (GW) detectors (comprising the two LIGO interferometers~\cite{TheLIGOScientific:2014jea} and the Virgo interferometer~\cite{TheVirgo:2014hva}) 
has completed three observing runs with $\sim$50 detections of
merging compact binaries~\cite{Abbott:2020niy}. 
% CHECK: Above, also add citations to IAS and AEI catalogs.
A fourth detector in Japan~\cite{KAGRA:rev:2020} is now being commissioned and is expected to join the global network in 2022.
A third LIGO detector with comparable astrophysical reach is being built in India~\cite{LIGOIndiaProposal:2011} and is
% The title of the paper says LIGO-India, hence we need to introduce it before transitioning to LIGO-Aundha:
expected to be operational during the latter part of this decade.
%Multiple detectors operating at different parts of the globe will provide several pairs of interferometers with longer %baselines and higher network SNRs than we have today. This will tremendously help improve the sky localization of GW events.
Several detectors operating in different parts of the globe provide multiple long baselines and an increased network SNR. These characteristics help improve the sky-localization of GW events, among other things~\cite{Ligo-india-Fairhurst-2014}.
Multiple detectors operating simultaneously will also improve the duty factor of the network leading to improvements in the detection rates.

In this paper we quantify the improvements arising due to the addition of a LIGO detector in India to the LIGO Global Network (LGN).
% new tekt - Rana
The global GW detector network will include, additionally, Virgo and KAGRA, further enhancing the improvements described herein. In this work, we choose to focus on the LGN to understand the improvement in the network during times when Virgo and KAGRA are not taking data.
We quantitatively describe how this leads to better astrophysical insights about the source properties and how that improves our ability to probe fundamental physics and cosmological models. We find that the addition of a new detector in India brings substantial benefits to the scientific capabilities of the LGN.

\subsection{Detectors}
\label{sec:detectors}
The LGN will consist of 3 interferometers in the upgraded configuration of Advanced LIGO (so-called A+)~\cite{aplus:2015}, with the third detector in Aundha, in the Hingoli district, in the eastern part of the state of Maharashtra, India. It is expected that the two LIGO detectors in the U.S.
% -- in Hanford, WA (H) and Livingston, LA (L) -- 
will be upgraded into this configuration in $\sim$2026 and that the detector in Aundha will come online soon after.
Following the existing naming convention~\footnote{where the detectors are named after the nearby town; LIGO--Hanford (H) and LIGO--Livingston (L)} the detector in India will be referred to as LIGO--Aundha (A). The LIGO Global Network with and without LIGO-Aundha will be denoted as AHL and HL, respectively.
Our studies below compare their performances, mainly related to the compact binary coalescence searches. Networks involving additional detectors will likely see further improvement in performances than what is found here. Moreover, the involvement of other detectors may reduce the impact of the improvements that the addition of LIGO-Aundha alone would bring. The broader study is, however, beyond the scope of this work. 
%in $\sim2027$.
%\begin{itemize}
%  \item we have 3 interferometers
%  \item they are in the A+ configuration
%  \item we expect Aundha to start taking data in 2027
%\end{itemize}

\begin{figure}
	\centering
	\includegraphics[width=\columnwidth]{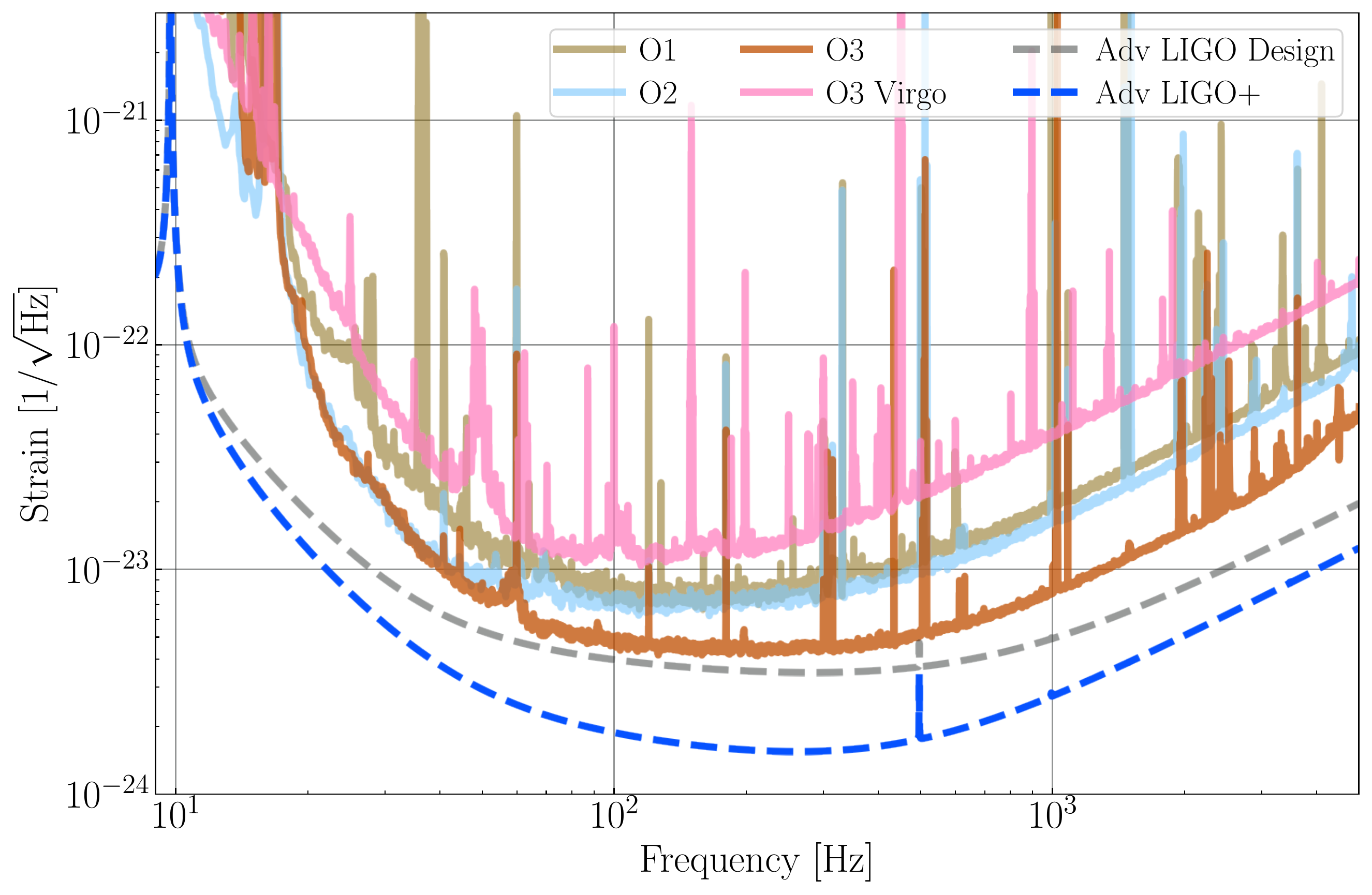}
	\caption{Strain noise spectral density of the LIGO Interferometers during the observing runs O1-O3. Also shown are the Virgo O3 noise, the Advanced LIGO design sensitivity, and the A+ sensitivity for the LIGO detectors Aundha, Hanford, and Livingston (labeled as ``Adv LIGO+").}
	\label{fig-noise}
\end{figure}

\subsection{Simulations}
\label{sec:sim}
%\cmt{This subsection is complete and ready for reading}

The compact binary coalescences (CBCs) observed in the GW window range in total mass from 3\,--\,150\,$M_\odot$. While most of the binary systems harbor primary objects with masses $< 45\,M_\odot$, a few systems have the primary heavier than $45 M_\odot$. The recently released gravitational-wave transient catalog (GWTC-2) considers several population mass distribution models to obtain the merger rates~\cite{Abbott:2020niy,Abbott:2020gyp}. For the simulations in this study, we use one of those mass models with the model parameters taken from the observed binary black-hole mergers~\cite{Abbott:2020gyp}. In this model, the primary mass follows a power-law distribution with some spectral index up to a certain maximum mass and a uniform Gaussian component with a finite width to account for high masses, together with a smoothing function at low masses to avoid a hard cut-off. The mass ratio follows a smoothed power-law distribution. We choose the median values of the hyper-parameters of these distributions inferred in~\cite{Abbott:2020gyp} for simulations.

%Astrophysical models suggest black holes can form in a binary system with isotropic spins in a dense environment 
Astrophysical models suggest that binary black holes with isotropically distributed component spins can form in dense environments,
such as globular clusters and galactic centers. 
At the same time, we expect black-hole spins to get aligned with the orbital angular momentum in isolated binaries~\cite{Kalogera:1999tq,Rodriguez:2016vmx}. The black hole spin distribution uses a model that is a mixture of both these possibilities~\cite{Abbott:2020gyp}. We use this model for drawing the spins of both the compact objects in a binary for our simulations. Besides, the binary sources are oriented uniformly and distributed uniformly over the sky and placed uniformly in co-moving volume up to a red-shift of 1.5 using the Planck 2015 cosmology~\cite{Planck15cosmology}.

The binary black hole (BBH) simulations described above are used in various studies below on quantifying the improvement in the performance of the network arising from its expansion to include LIGO-Aundha. These include a discussion of BBH detection rates in~\Cref{sec:rates} and quantifying the improvement in the estimation of binary parameters in~\Cref{sec:PE}. In~\Cref{sec:sky_localization}, we discuss the possibility of sending early warning alerts to electromagnetic and particle observatories before the epoch of binary coalescence. We make projections in~\Cref{sec:tog} on how a detected BBH population can be used to place observational bounds on deviations from General Relativity (GR).

%The results in this paper are derived from analyses performed on a set of injected compact binary signals. The properties of these injections such as mass and spin distributions follow the astrophysical properties inferred from the
%GWTC-2 catalog \Rana{not the O3a catalog?}.
%For binary black hole population, the salient features are listed below:
%\begin{itemize}
%    \item Mass distributions
%    \item Spin distribnution
%    \item $D_L$ distribution (or redshift)
%\end{itemize}

\section{CBC Detection rates }
\label{sec:rates}

Coalescing compact binaries involving neutron stars and black holes are, so far, the only GW sources detected in past GW observing runs~\cite{LIGOScientific:2018mvr, Abbott:2020niy}. 
The inclusion of LIGO-Aundha in the LGN will boost the rate at which we detect such binaries. This enhancement will arise owing to improved sky-coverage, distance reach, and baseline duty factor, which is the effective observation period of a detector network. In this section, we quantitatively assess the improvement in the CBC detection rate ($R_{\rm det}$) for AHL {\it vis \`{a} vis} the HL network. 

We focus here on the stellar-mass BBHs, which are the main contributor to the menagerie of signals observed by LIGO-Virgo so far. Our analysis can be straightforwardly extended to classes of CBC sources that involve neutron stars. 
%We perform simulations of gravitational-wave signals from these systems and compute the signal-to-noise ratio (SNR) for the AHL using A+ PSD. The masses and spins of these CBCs follow astrophysical distributions consistent with the properties inferred from the GWTC-2 catalog~ \cite{Abbott:2020niy, Abbott:2020gyp}. Moreover, we take the sources to be distributed uniformly in comoving volume up to redshift of 1.5, following standard cosmology.

% Though there are different classes of CBC sources such as BNS, BBH, NSBH and IMBHB which all might have different formation %channels, we restrict our study to only BBH events, since the improvement factor can be generalized to other classes as well. 
 
For an astrophysical population of BBHs with a comoving constant merger-rate density $r_{\rm merg}$ in units of ${\rm Gpc}^{-3} {\rm yr}^{-1}$, the detection rate (per year) is
given by
$R_{\rm det} = r_{\rm merg} \times \langle VT \rangle \,$,
where $\langle VT \rangle$ is the population-marginalized detection volume averaged over the period of observation for any given detector network (for more details, see \cite{ Abbott:2020niy, LIGOScientific:2018jsj} and the references therein). 
The assumption that $r_{\rm merg}$ is  non-evolving {w.r.t.} redshift is a simplified assumption and hence could affect the rates we reported in this paper, however it has negligible impact on the rates comparison between two networks which is the goal of this study.  The factor $\langle VT \rangle$ crucially depends on the number of detectors, their sensitivity as well as the search methodologies and their ability to treat the non-Gaussian noisy transients in the multi-detector data. 
%Continuous efforts are made on improving the sensitivity of the detectors as well as inclusion of novel ideas to combat noisy transients and hence improve the detection algorithms. 

%We expect the LIGO-Aundha to contribute in increased detection rates in terms of increased instantaneous detection volume (V) as well as %improved duty cycle in terms of the effective observation period $T$ (also known as the effective duty cycle or the network duty cycle).  

\subsection{Detection criteria}

We simulate the noise in any detector as Gaussian, with a vanishing mean, and uncorrelated with the noise in any other detector.~\footnote{This is a simplification since real detector noise contains non-Gaussian transients, which contribute to the background rate. Still modeling detector noise as Gaussian is useful. As has been demonstrated in multiple LIGO-Virgo CBC and detector characterization papers~\cite{LIGOScientific:2019hgc,LIGO:2021ppb,Bose:2016sqv,Bose:2016jeo}, glitch classification and mitigation techniques have achieved some degree of success in cleaning the background to make it largely Gaussian-like. Simulation studies, like ours, are not the first ones, and are useful also for providing targets and benchmarks for those data quality/cleaning efforts. It is for these reasons Gaussian studies remain relevant.}  
%assuming that the non-Gaussian background will contribute as in previous runs\cite{detchar}}.  
Then the network coherent SNR-squared is the sum of the SNR-squared of signals in the individual detectors \cite{Finn:2000hj,Pai:2000zt}.   
%discuss how to choose a detection criterion that can represent the detections of GW signals with a multi-detector network. 
Below we discuss two alternative criteria for assessing whether a signal can be considered as detected by a network (similar considerations are made in \cite{OSD}):

\begin{enumerate}
\item {\bf Coherent network SNR criterion}:  For an $N$-detector network, this criterion is $\sqrt{\sum_{k=1}^{N} \rho_k^2} \geq \rho^{\rm net}_{\rm thresh}$, where $\rho_k$ is the SNR at the $k^{th}$ detector. Here we set the threshold of $\rho^{\rm net}_{\rm thresh}$ to a value that keeps the false-alarm probability associated with it low enough to make a confident detection case. 
%The  depends on the data analysis techniques used by the pipelines. 

\item  {\bf  Multi-detector coincidence criterion}: $\sqrt{\sum_{k=1}^{N} \rho_k^2} \geq \rho^{\rm net}_{\rm thresh}$ and $\rho_k > 4$ for at least two of the $N$ detectors.

\end{enumerate}

For the LGN studied here we set $\rho^{\rm net}_{\rm thresh} = 12$, which is conservative in the sense that there have been detections with two or three detectors with network SNR below 12. %(thereby, allowing for computational tractability of follow-up analysis required for establishing astrophysical significance) nor too low (hence, keeping the false-dismissal rate of signals low).
We present search performance metrics for both criteria below.

Arguably, the simplest way to identify interesting detection candidates is to apply the first criterion. Its biggest advantage is that it allows for picking up sources that are loud enough in one detector but weak in the others, e.g., if located in their blind-spots. This can happen since no two detectors have the same orientation. 
%%is useful to {\color{red} [Ambiguous, Saleem:] obtain the distance reach}, 
Nevertheless, this criterion has a few limitations: For instance, a loud noise-transient in a single detector ({\it e.g,} non-Gaussian glitches) can give rise to a trigger that satisfies this network criterion and, therefore, gets misclassified as a detection candidate. On the other hand, if one requires that at least two detectors record a high enough SNR, such as what the second criterion above employs, then the false-alarm rate reduces significantly (such as by mitigating the effects of non-Gaussian glitches), albeit by sacrificing some degree of sky coverage (figure \ref{fig-noise}). 

%the detectability of sources from certain parts of the network sky.  
%%This will reduce the number of detections compared to the former criterion, but the detections will be of higher detection significance, which would be important for probing fundamental physics without biased inferences. 

\subsection{Improvement in the effective duty-factor of a network}

\begin{figure}
    \centering    
    \includegraphics[width=0.48\textwidth]{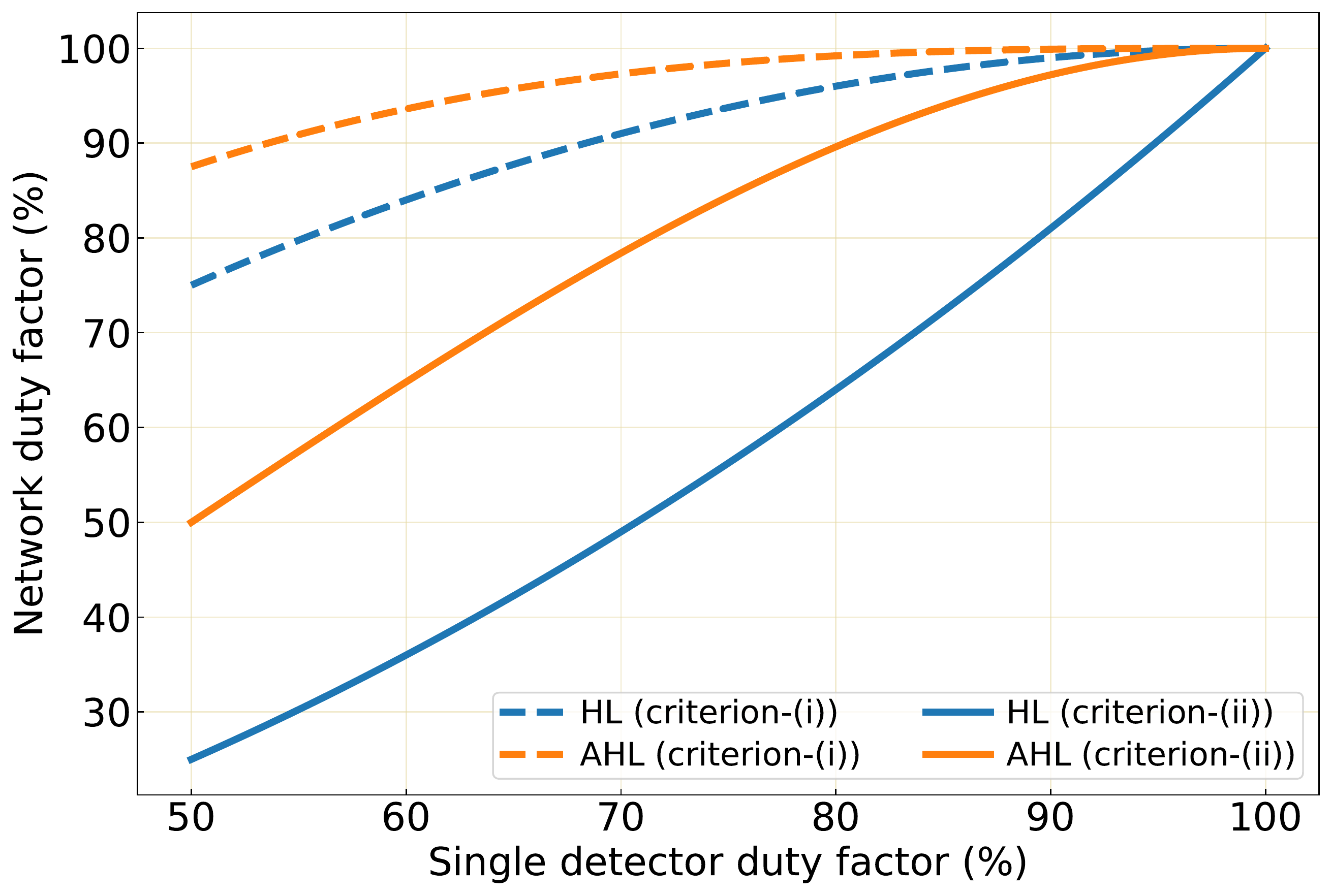} 	
    \includegraphics[width=0.48\textwidth]{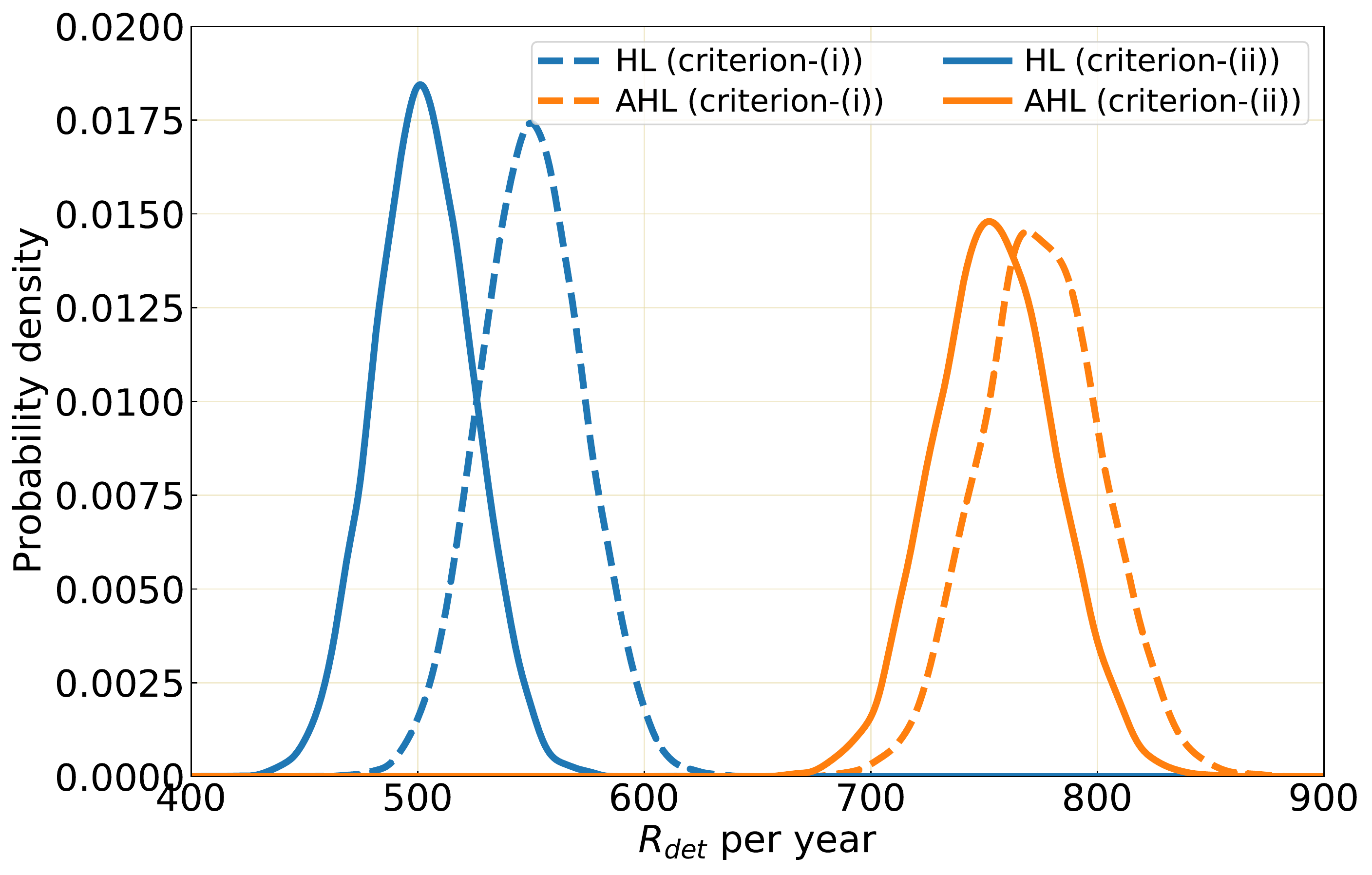}
    \caption{[Left] Network duty factors of the HL and AHL networks as functions of the single-detector duty factor. [Right] The distribution of the detection rates of stellar-mass binary black-hole detection rates for the same two networks using the GWTC-2 population models.}
    \label{fig:dc-rates}
\end{figure}

Duty factor of a detector (network) is defined as the fraction of clock time for which the detector (network) acquires science quality data.  Assuming that each detector in the network has a duty factor of $d_f$, one can analytically compute the effective duty factor $\textrm{d}_{f\>\textrm{eff}}$ for each multi-detector network. For the multi-detector coincidence criterion, $\textrm{d}_{f\>\textrm{eff}}$ is the fraction of the observation period during which at least two of the detectors are simultaneously collecting science-quality data while for the network SNR criterion, it is the fraction of observing period when at least one of the detectors is observing in science mode. 
%Alternatively, the definition of effective duty factor for the first (second) criterion can be expressed as the probability that at least one (two) detectors are giving useful data at any instant of time. Hence, 
For an $N$-detector network, the effective duty factor is
\begin{equation}
\textrm{d}_{f\>\textrm{eff}} = \sum_{k=N_{\rm min}}^{N} {}^{N}C_{k} \, d_f^k \, (1-d_f)^{N-k} \,,
\end{equation}
where the summation runs from $k=N_{min}$ to $k=N$ with $N_{\rm min}$ being the minimum number of detectors required by the coincidence criterion.  Specifically, we have $N_{min}=1$ for criterion-(i) and $N_{min} = 2$ for criterion-(ii). The combinatorics symbol 
%$C$ denotes the standard notation in the combinational problem such that 
${}^{N}C_{k}$ denotes the number of possible unique $k$-detector combinations one can form in a network of $N$ detectors. 
\Cref{fig:dc-rates} shows how the effective duty factor improves with the addition of LIGO-Aundha. Assuming 90\% single-detector duty-factor, the AHL duty-factor 
%{\color{red} [Saleem: (the figure is missing )]} 
gets boosted by a factor of $\sim 1.2$ compared to the HL network if one follows criterion-(ii),
while the improvement is only one per-cent under criterion-(i); see, e.g., \Cref{fig:dc-rates}.~\footnote{Here it is assumed that the unlocked time-stretches are randomly and uniformly 
distributed over the full observation period, which excludes any stretches of time scheduled 
for concurrent downtime for all detectors.}

% Plot script
% https://git.ligo.org/ligo-india/science_case/-/blob/master/scripts/rates_plots_and_numbers.py

\subsection{Detection rates}

We perform extensive simulations to estimate the detection rates of merging compact binaries for the HL and AHL configurations. 
%We compute the detection volume, $\langle VT \rangle$ in Eq.~\ref{eq:Rdet} which depends on the network configuration, detector sensitivity and the distribution of the source population. We simulate a stellar binary black hole merger population that follows the population properties as inferred from the GWTC-2 catalog CITE[GWTC-2:2020]. Using the Monte-Carlo simulatation, we estimate $\langle VT \rangle$ as,
%\begin{equation}
%    \langle VT \rangle = %\left(\frac{N_{\rm det}}{N_{max}}\right) \times V_{max}.
%\end{equation}
%where $N_{max}$ is the total number of sources in the simulated population distributed in a spherical volume $V_{max}$, and $N_{\rm det}$ is the number of detected binary mergers under a given criterion. 
The published detections of binary black hole mergers provide an up to date median BBH merger rate density of $r_{\rm merg} = 23$ Gpc$^{-3}$ yr$^{-1}$ \cite{Abbott:2020gyp}. With this rate density and a uniform source distribution in the comoving volume, we populate $\sim 8684$ sources up to a redshift of 1.5.~\footnote{It is a somewhat arbitrary choice that we truncate the population at a maximum redshift of 1.5. However, this is motivated by the fact that at higher redshifts, the actual comoving rate density could significantly be different from the merger rate density at $z=0$ (the one we assumed in this study) due to the star formation rate as well as the distribution of delay time between the formation and the coalescence of the binary.} We perform 8000 batches of simulations, with each batch containing 8684 sources with the mass and spin distributions following the ones detailed in \Cref{sec:sim}. We further distribute the sources uniformly over the sky with the binary orientation distributed uniformly. 
%We show that with the AHL configuration, the network SNR on an average, increases by a factor 1.4 with respect to the HL network configuration. 
For all the sources, we apply the detection criteria (i) and (ii) and obtain the detection rates ($R_{\rm det}$). The right panel of \Cref{fig:dc-rates} provides the distribution of the estimated $R_{\rm det}$.

\begin{table}\centering 
\begin{tabular}{|l|c|c|c|}
\hline        
Network  &  Criterion (i) 	             &  Criterion (ii)  \\
\hline
HL	     &  $550.0^{+30.0}_{-29.0}$ 	 &  $502.0^{+29.0}_{-27.0}$ 	  \\
AHL		 &  $775.0^{+35.0}_{-35.0}$ 	 &  $754.0^{+35.0}_{-33.0}$	  \\
\hline
\end{tabular}
\caption {Detection rates (in ${\rm yr}^{-1}$) of stellar-mass binary black holes in HL and AHL networks, assuming A+ sensitivity and a duty factor of 90\% for every detector.}            
\label{tab:bbh-rates}
\end{table}

\Cref{tab:bbh-rates} provides the detection rate estimates of BBH for HL and AHL network configurations assuming 90\% single-detector duty factor. Compared to HL, the detection rate in AHL increases by $41 \%$ and $50 \%$ for criteria (i) and (ii), respectively. Besides the duty factor, the sky coverage of the two networks determines their detection rates. 
%% wording is confusing: and hence subject to vary, though we have assumed all the three detectors at the same sensitivity. 
Since the coincidence criterion-(ii) exhibits a preference for shortlisting highly significant events, one can expect that with LIGO-Aundha one will see a perceptible increase in such events under that criterion.

%\begin{figure}
 %   \centering    
%  \includegraphics[width=0.48\textwidth]{../figures/bbhDLmax.pdf}
% \caption{The population distribution of the luminosity distance to the farthest BBH merger that can be detected with HL and AHL networks.}
%    \label{fig:DLmaxBBH}
%\end{figure}

%\subsection(Farthest distance observed)
%The simulations used for computing the rates also allow us to compute the farthest detectable BBH merger by picking the highest $D_L$ value among the detected sources in each of the 8000 set of simulations. In \Cref{fig:DLmaxBBH} we show the distribution of the farthest detectable BBH merger with AHL and HL using the two detection criterion. With the multi-detector coincidence criterion, we find that the farthest BBH that would be detected with HL would be in the range $xxxx - xxxx$ Mpc while with AHL, this would be in the range $xxxx - xxxx$ Mpc. Note that these estimates are subject to the underlying assumptions of the population models described above. Further, we have computed this for the entire population distribution rather than considering any specific mass bin. Therefore, the BBH mergers detectable from these ranges will be from the higher end of the mass spectrum while those from the lower end will be detectable up to much lesser distances.     

\section{Parameter estimation} \label{sec:PE} %(we removed LGN from title)
%\cmt{Assigned to: Saleem, Gayathri and Archana. This section is complete and ready for reading}

With the expansion of the LIGO global network and the consequent enhancement in the signal-to-noise ratios of the CBC detections and mitigation of parameter degeneracies, one would anticipate improvements in the astrophysical parameter estimation. In this section, we employ CBC signal simulations to obtain quantitative support for this expectation.
%demonstrate the expected improvements in the estimated binary parameters, with the help of simulated CBC signals.

For a BBH system in a circular orbit, the gravitational-wave signal is characterized by component masses ($m_1, m_2$), component spins ($\vec{S_1}, \vec{S_2}$), the luminosity distance ($D_L$), orbital inclination angle ($\iota$), polarisation angle ($\psi$), sky-position angles ($\alpha, \delta$) and the coalescence time and phase ($t_c, \phi_c$). For a binary neutron star (BNS) system, we require at least two additional parameters in the form of component tidal deformability parameters ($\Lambda_{1}, \Lambda_{2}$).
%The sensitive frequency range is high for the ground-based detectors, and thus the gravitational-wave signals observed in the band are expected to be insensitive to their initial eccentricities.

The CBC signal's multi-dimensional parameter space harbors correlations and degeneracies among different parameter pairs, contributing to the uncertainties in the measurements of the individual parameters. For several of these parameters, the error-bar scales inversely with the signal-to-noise ratio (for loud signals)~\cite{Cutler:1994ys}. While this holds particularly well for the intrinsic binary parameters, such as component masses and spins, the aforementioned degeneracies among some pairs, e.g., (i) the sky-location angles $\alpha$ and $\delta$ and (ii) $d_L$ and $\iota$,
can not often be removed despite high SNR. The expansion of LGN with LIGO-Aundha, in addition to increasing the SNR, will enhance parameter estimation accuracy by providing an independent observation of the source that can significantly reduce the degeneracies among some of the parameters.

In Sec.~\ref{subsec:bbh_pe} we focus on general parameter estimation for select BBH events, and in Sec.~\ref{sec:eos_pe} we present the primary results of masses and tidal effects in BNS systems.

\subsection{Improvement in errors for binary black hole events}
\label{subsec:bbh_pe}

For this study, we simulated binary black hole signals modeled after two of the observed binary black holes, namely, (i) the loudest BBH,  GW150914~\cite{Abbott:2016blz} and (ii) the most massive BBH, GW190521~\cite{Abbott:2020tfl}. In fact, GW150914 is the first binary black hole merger observed by two LIGO detectors and the loudest event so far, with a coherent SNR of 24. The observed component masses were 36\,$M_\odot$ and \,29\,$M_\odot$ with a remnant BH of 62\,$M_\odot$, and the event was located at a luminosity distance of 450\,Mpc. It was localized in a huge sky-patch, spanning $590$ sq.~degs.

GW190521 is the most massive and among the farthest (5\,Gpc) binary black hole mergers observed so far. Its component masses are 85\,$M_\odot$ and 66\,$M_\odot$. The remnant was estimated to have a mass of $142\,M_\odot$. This is the first intermediate-mass BH candidate observed in the gravitational-wave window.

For our two simulations, the injected values of the key parameters are listed in Table~\ref{tab:pe} where we choose the masses and spins to be identical to those inferred for GW150914 and GW190521.
From our 8000 batches of BBH simulations described in Sec.~\ref{sec:rates}, it was found that the population-averaged ratio of SNR at AHL to the SNR at HL lies in the range of 1.3\,--\,1.4.  We choose the injected sky positions in such a way that the SNR at AHL is $\sim1.4$ times the SNR at the HL so that it resembles the average behaviour of SNR improvement.
% We inject the sources at an average sky-location such that the SNR of the event with LGN is improves \textcolor{red}{$\sim 1.4$}  times that of the LIGO-US - a number which is approximately equal to the sky-averaged SNR ratio between LGN and LIGO-US. This is to ensure that we do not over-estimate or underestimate the PE improvements by placing sources either at exceptionally bright or exceptionally dark spots of the LIGO-US network.

The run-of-the-mill Bayesian parameter estimation approach assumes stationary Gaussian detector noise and a reliable, faithful Einstein's GR signal model for the GW signal from the compact binary merger. An up to date suite of models for complete CBC waveforms constructed by combining various approaches include phenomenological models, such as IMRPhenom models~\cite{WF-IMR-ajith2011}, the effective one-body EOBNR waveforms that use inputs from numerical relativity~\cite{Buonanno:1998gg,Buonanno:2000ef,Barausse:2009xi}, and the NRSurrogates waveforms derived from numerical relativity simulations~\cite{Field:2013cfa,Blackman:2015pia,Varma:2018mmi}. In our analysis, we use the \textit{IMRPhenomPv2}~\cite{Hannam:2013oca,Schmidt:2014iyl,Husa:2015iqa,Khan:2015jqa,Bohe:PPv2} waveform model for both injections as well as recovery. We use the \texttt{Bilby}~\cite{bilby} software package, with its in-built sampler \texttt{dynesty}, to perform the parameter estimation. We perform this analysis with \textit{zero-noise} signal injections~\footnote{A zero-noise signal injection refers to data that has only a simulated GW signal and no added noise.} and the likelihood computed using the A+ PSD. 

We tabulate the results in terms of improvement in the 90\% credible intervals on various astrophysical parameters in Table~\ref{tab:pe} and present pictorially in Fig.~\ref{fig:PEparam} the posterior probability contours (at 90\%, and 68\%  credible levels). In that figure, the left and right panels depict the results for the GW150914-  and GW190521-like injections, respectively.

\begin{table*}[!h]

\begin{center}
\begin{tabular}{ |c | c | c | c | c |}
\hline
\multicolumn{1}{| c |}{\textbf{Parameter}}  & \multicolumn{2}{ c |}{\textbf{GW150914-like}} & \multicolumn{2}{ c |}{\textbf{GW190521-like}} \\
%---------------
\cline{2-5}%\cline{6-8} \\
& Injected   & Improvement & Injected & Improvement\\
\hline\hline
Chirpmass $(M_\odot) $ & 28.1  & 33\%  & 64.6  & 39\% \\
Total mass $(M_\odot)$ & 65.0  & 33\%  & 149.6 & 40\% \\
$D_L$ in Gpc           & 2.5   & 35\%  & 5.3   & 36\% \\
$\iota$ in deg.  & 45   & 27\%   & 45   & 70\%  \\
Sky localization in deg$^2$.  &    & 92\%   &    & 96\%  \\
%$RA$ in deg.           & 212   & 4 & 212   & 22  \\
%$Dec$ in deg.          & 166   & 13 & 166   & 34 \\
%$\chi_{eff}$           & 0.2   & 28\%  & 0.1   & 25\% \\
%$\chi_p$               & 0.2   & 30\%  & 0.7   & 14\% \\
\hline%\hline
\end{tabular}
\end{center}
%\vspace{-0.5cm}
\caption{Parameter estimation improvement in 90\% credible intervals in expanding the network from HL to AHL: We use BBH signals modelled after GW150914 and GW190521 and estimate the improvement in sky-localization, luminosity distance, binary inclination, masses and spins. The imporvemnt for a parameter $X$ is defined as $((\Delta X_{AHL}- \Delta X_{HL})/\Delta X_{HL}) \times 100 $ where $\Delta X$ is the 90\% credible error bar.}
% the data in this table is generated by https://git.ligo.org/ligo-india/science_case/-/blob/master/scripts/pe_table_updated.ipynb
\label{tab:pe}
\end{table*}

% $\theta_{jn}$ is now defined as $\iota$ everywhere.

\subsubsection{Sky-localization:} The detector pair comprising LIGO-Aundha and LIGO-Livingston provides the longest baseline amongst all pairs of existing / in-construction detectors. This improves the precision with which sources can be localized in the sky. For the GW150914-like injection, the 90\% credible 2-D localization area is $\sim 114$ deg$^2$ which improves to $\sim 9$ deg$^2$ with AHL. This amounts to $~92\%$ reduction in the localization uncertainty. For GW190521-like injection, we find a $\sim 96$\% reduction, with the respective localization area for HL and AHL configurations being $\sim 971$ deg$^2$ and $\sim 35$ deg$^2$.
%the RA and Dec errors improve by 38\% and 58\%, respectively; while for the GW190521-like injection the respective improvements are 73\% and 75\%.
More discussion on the sky-localization can be found in Sec.~\ref{sec:sky_localization} and the reader may also refer to earlier studies on localization, e.g., Refs.~\cite{Pankow:2019oxl,Ajith:2009fz} and the references therein. 

\subsubsection{Luminosity distance and inclination angle:} The three-detector configuration plays a crucial role in breaking the degeneracy between the luminosity distance $D_L$ and the inclination angle $\iota$. For the GW150914-like system, the errors in $D_L$ and the inclination shrink by 35\% and 27\%, respectively, for AHL relative to HL. Similarly, for the GW190521-like injection the error reduction in the same parameters is 36\% and 70\%, respectively. The improved distance estimates also benefit from the reduced 2D sky-localization of the source by the AHL network since, aided by an improved network SNR, it helps break the degeneracy between distance and sky position. This will have direct implications in the measurements of cosmological parameters~\cite{LV-H0-2019,LV-H0-2017,Schutz:1986gp,DelPozzo:2012zz,Nair:2018ign,Chen:2017rfc}. We also find significant improvement in the inclination angle measurement of the binary ($\iota$) which is partly due to the resolution of the distance-inclination degeneracy. Though this analysis has been performed on binary black hole mergers, similar improvements are expected in the inclination angles of binary neutron stars and neutron star-black hole mergers as well~\cite{Rodriguez:2013oaa} which are favourite candidates to have associated EM counterparts. Accurate knowledge of the inclination angle is key in doing multimessenger astronomy, for making predictions on the possible EM counterparts and in understanding the physical process that drives the EM counterparts~\cite{Arun:2014ysa,Saleem:2019,Saleem:2020gw190425}.
Further, improved precision in the binary inclination helps to probe the gravitational-wave polarisation of the signal. This improvement directly impacts probing alternative theories of gravity with  gravitational wave signals. In Sec.~\ref{sec:pol} we discuss how polarisation measurements benefit from the expansion of LGN.

%\subsubsection{Detector-frame masses}: For GW150914-like injection, the estimated detector-frame masses with LIGO-US network are $m_1 = 53.154^{+7.736}_{-5.156}$ and $m_2 = 39.846^{+4.447}_{-4.295}$ while the same with LGN network are $m_1 = 51.046^{+4.291}_{-2.944}$ and $m_2 = 42.038^{+3.08}_{-3.764}$, with the latter having error bars improved by 44\% and 22\% respectively. For GW190521-like injection, we get $m_1 = 157.72^{+19.65}_{-23.01}$ and $m_2 = 104.25^{+27.43}_{-15.14}$ for LIGO-US and  $m_1 = 141.99^{+14.83}_{-6.28}$ and $m_2 = 72.8^ {5.4}_{9.1}$ for LGN, with the improvements being 51\% and 37\% respectively. These improvements may be attributed to the improved SNR between the two networks.  \cmt{Archana: Saleem, do we need detector frame masses? They are not physical right?}

\subsubsection{Source masses:}
\begin{figure}[h]
	%\hspace{-1cm}
	\includegraphics[scale=0.32]{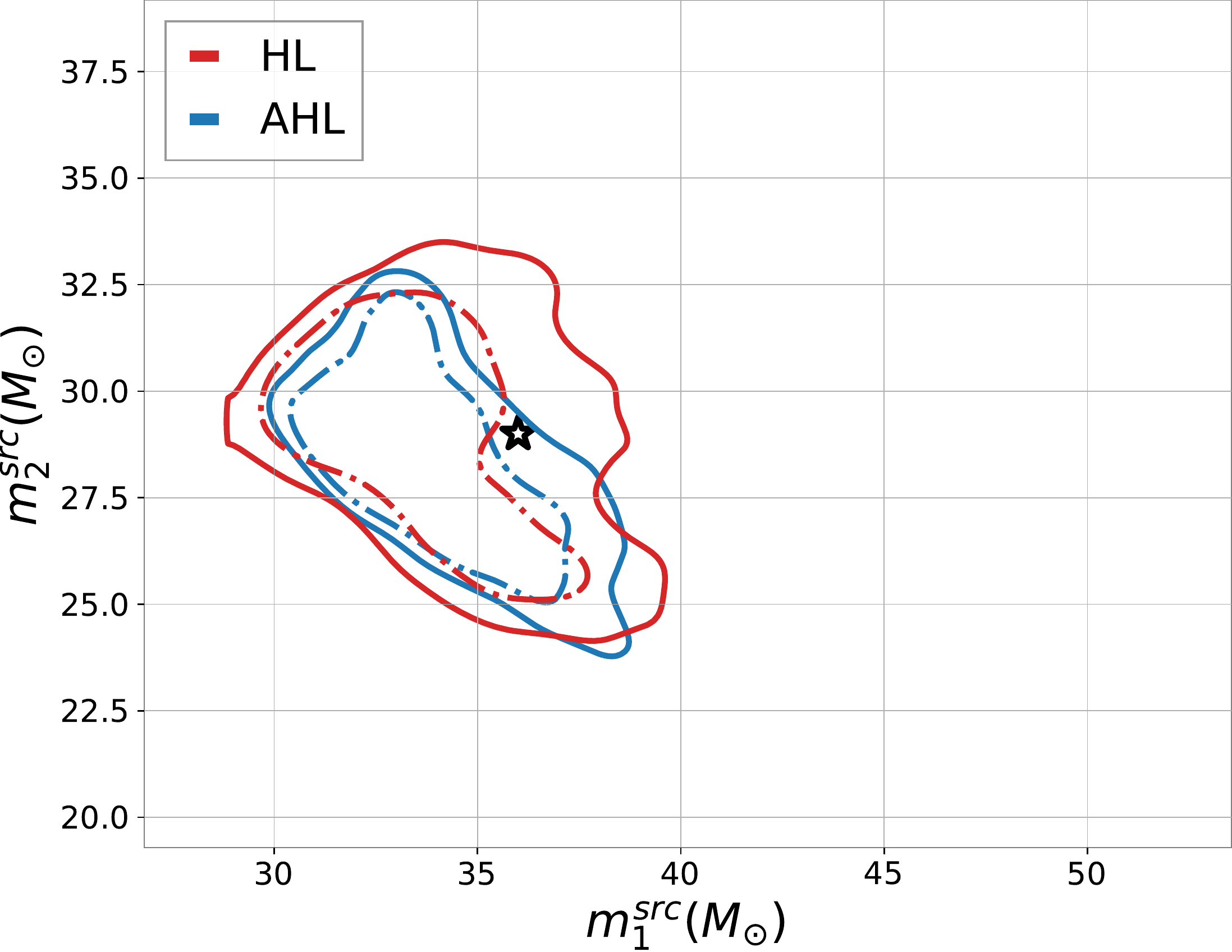}
	\includegraphics[scale=0.32]{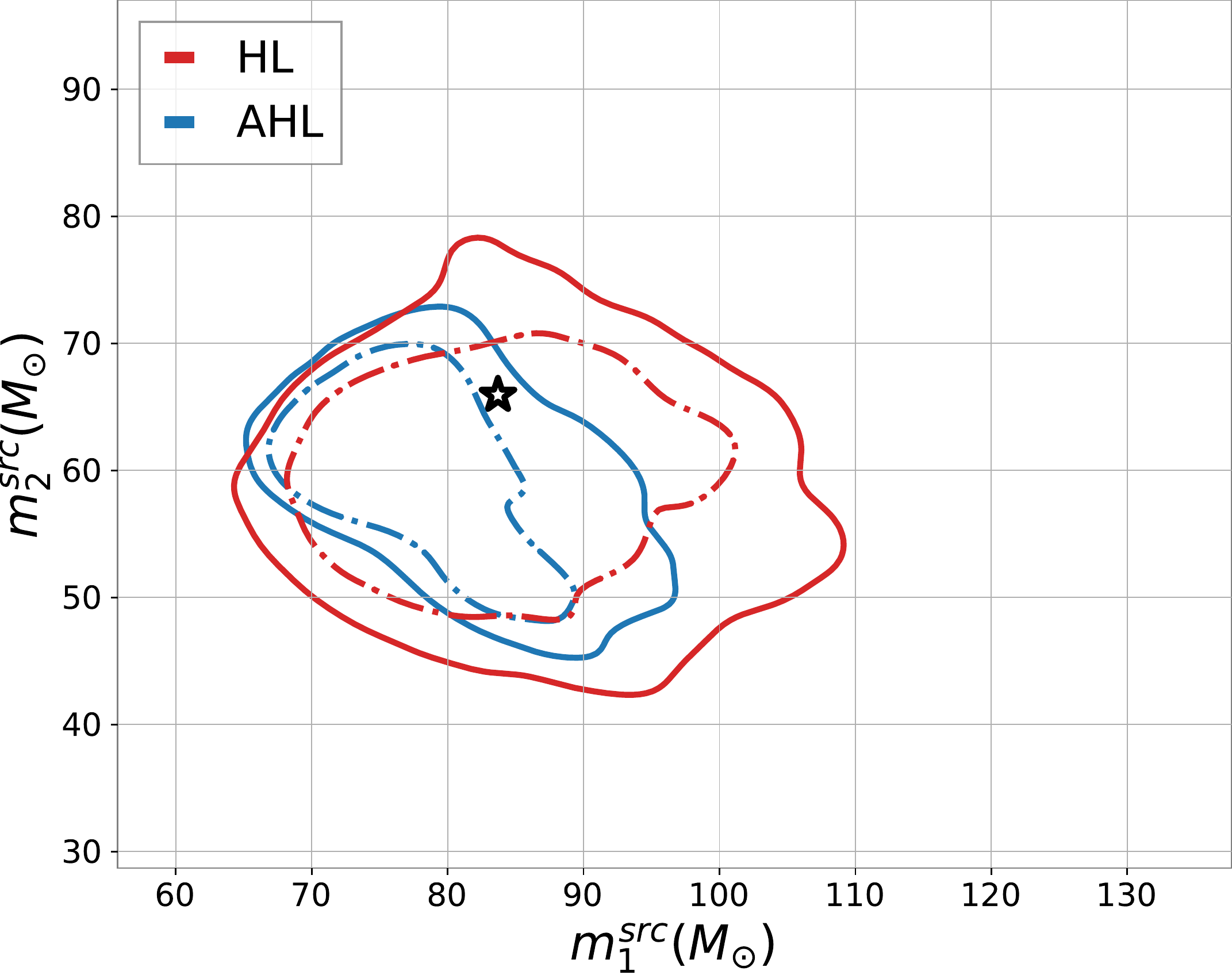}
	\includegraphics[scale=0.33]{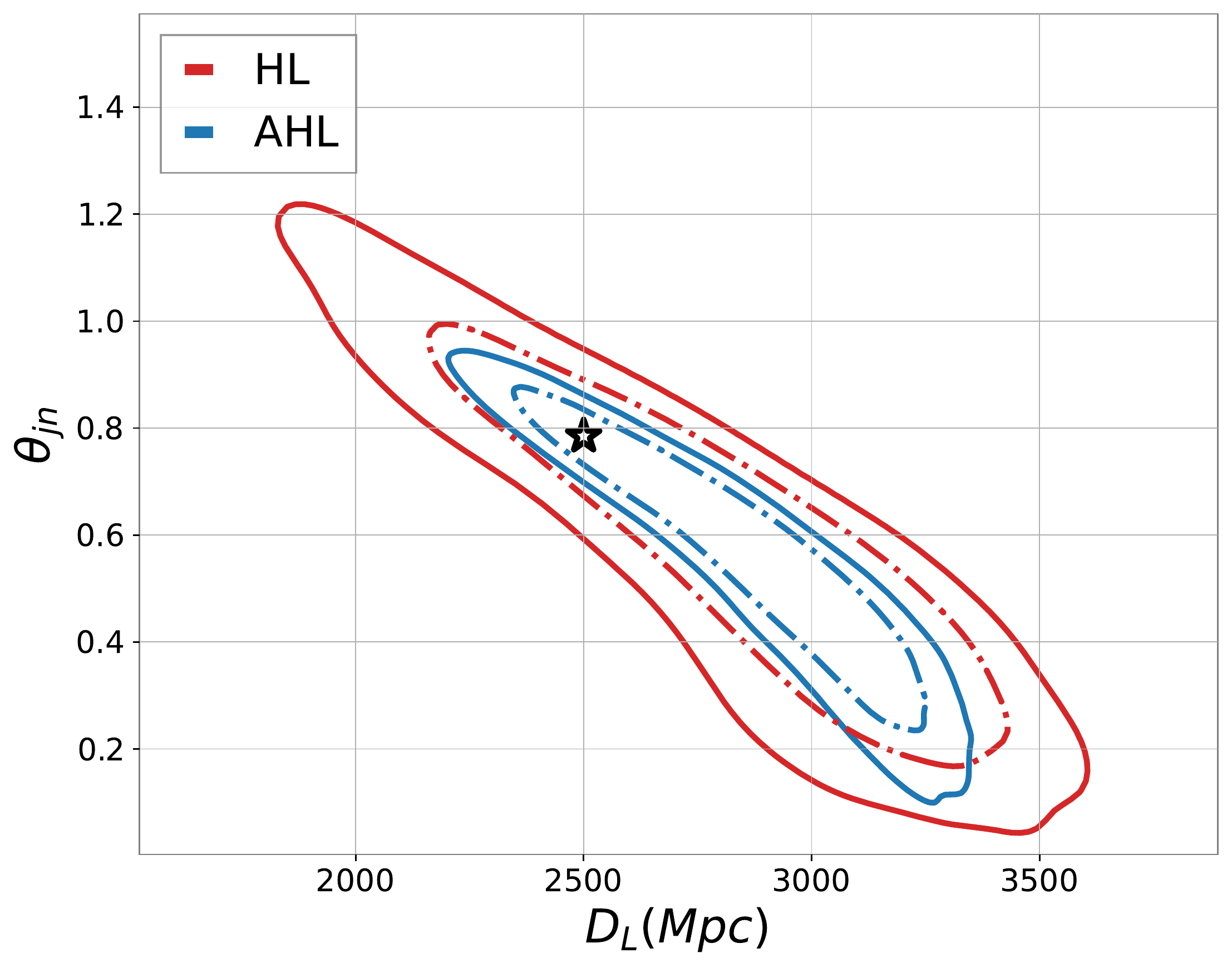}
	\hspace{1.3cm}
	\includegraphics[scale=0.33]{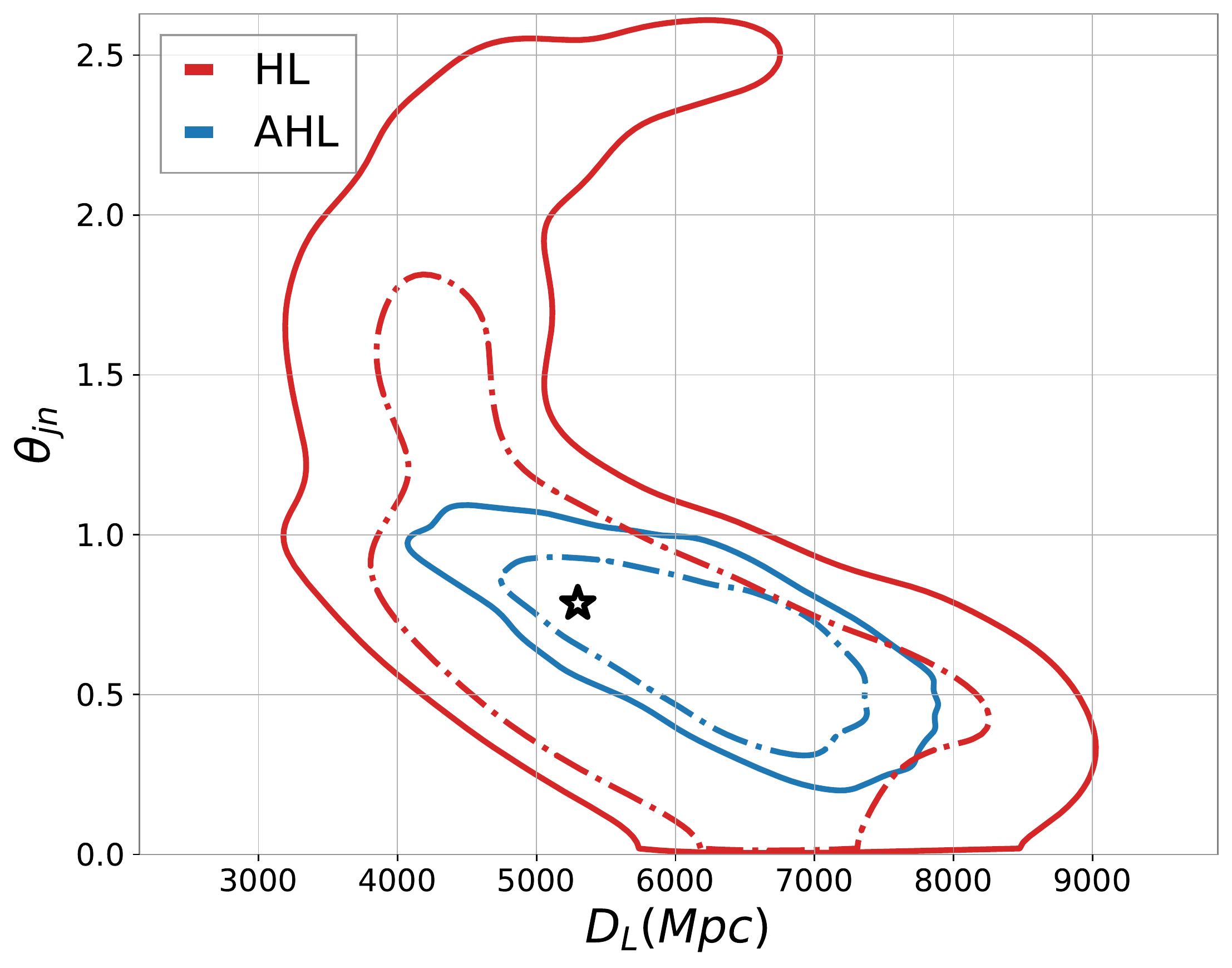}
	\includegraphics[scale=0.33]{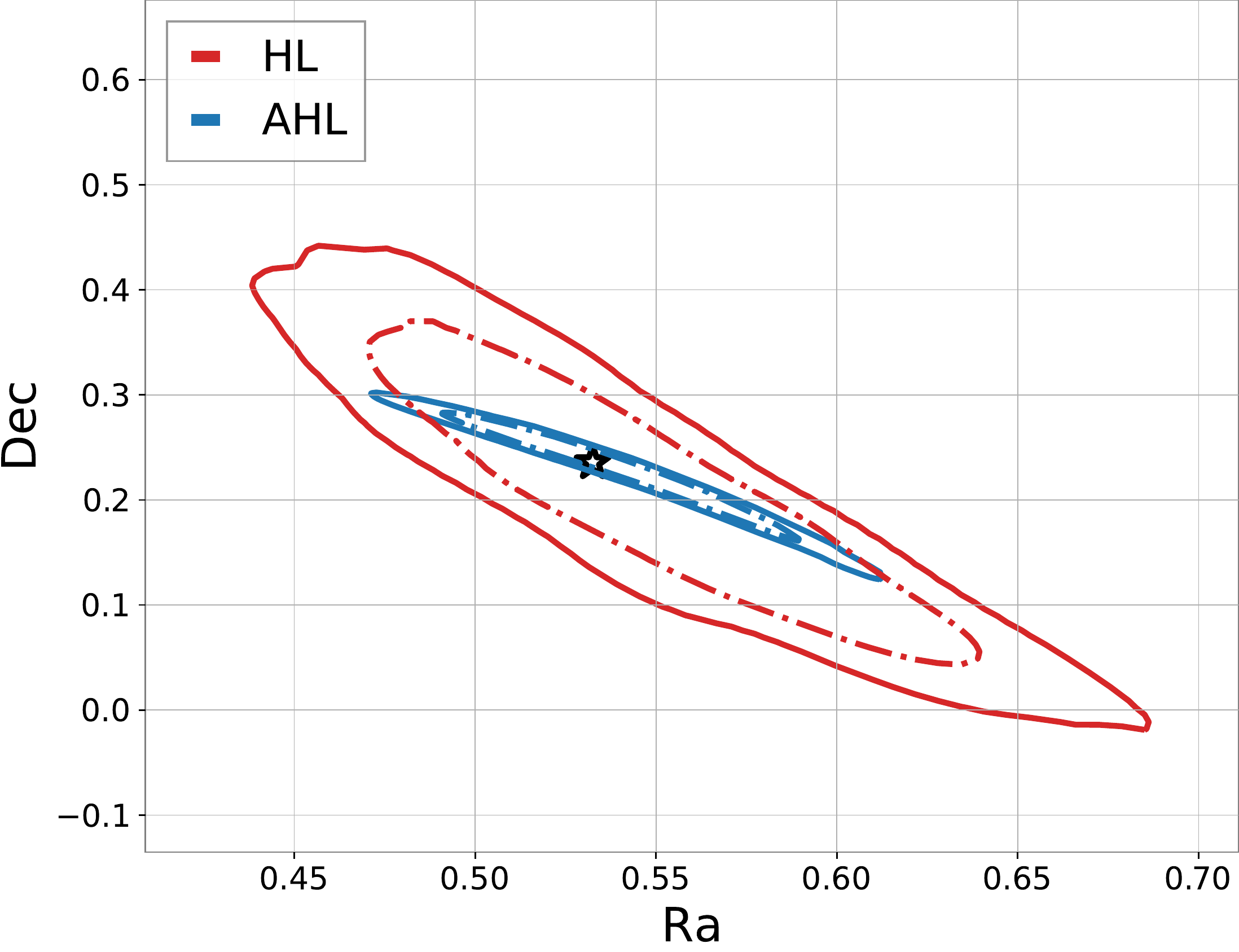}
	\hspace{1.3cm}
	\includegraphics[scale=0.33]{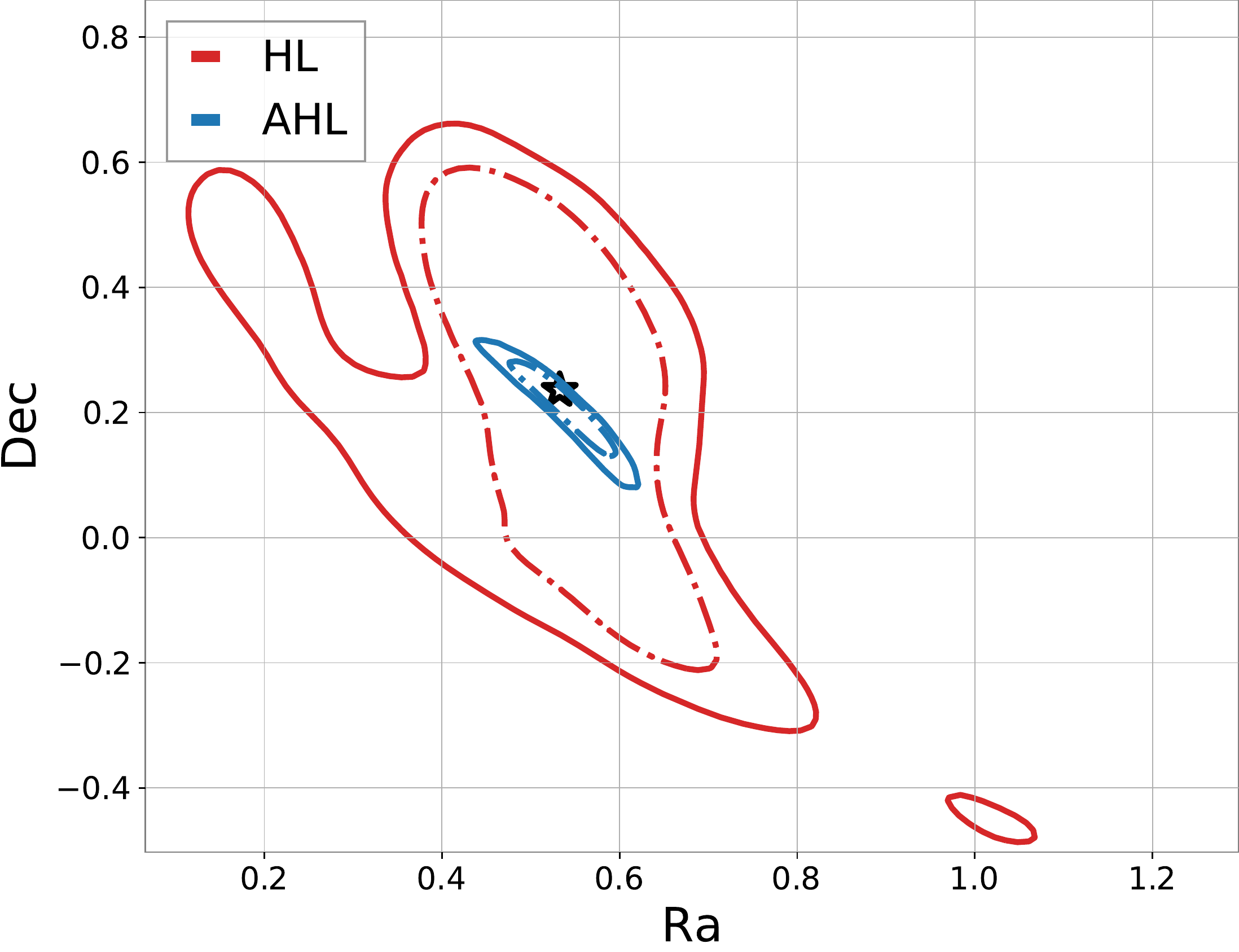}
        \caption{Posterior distributions of certain parameters for GW150914-like (left) and GW190521-like (right) simulated signals in the HL and AHL networks:  Top, middle and bottom  panels correspond to the parameters $m_1^{src} - m_2^{src} $, $D_L-\iota$ and $RA-Dec$, respectively.  The true values are shown by a black star.
          The 95\% and 65\% confidence intervals are shown by solid and dash-dotted lines, respectively.}
% script to produce scripts/pe_plots_GW150914.ipynb and scripts/pe_plots_GW190521.ipynb
\label{fig:PEparam}
\end{figure}

Source-frame masses are defined as the detector-frame masses divided by a factor $(1+z)$, where $z$ is the source redshift, which in turn can reveal the source luminosity distance given a cosmological model. Therefore, the measurement of source-frame masses benefits from both the improved SNR and the improved luminosity distance measurement. For the GW150914-like injection, the errors in both the source-frame chirpmass and total mass improve  by $\sim33\%$. Similarly, for the GW190521-like injection, these improvements are 39\% and  40\%, respectively. See Fig.~\ref{fig:PEparam} for the $m_1 - m_2$ contour plots for both the events. Accurate knowledge of the intrinsic source parameters helps in the population synthesis studies of compact binary mergers and obtain constraints on the merger rate density~\cite{Abbott:2020gyp}. %The spin parameters, as shown in the table, show improvements which are broadly consistent with what one would expect from the increased SNR\footnote{With the SNR at AHL being 1.4 times the SNR at HL, and assuming the error-bar scales as the inverse of SNR, the size of the error-bar is expected to reduce by 29\%.}.

\subsection{Improved measurements of matter effects: source classification and BNS properties:}
\label{sec:eos_pe}
%\cmt{Assigned to: Arunava} 

Binary neutron stars are characterized by the masses ($m_{1}$, $m_{2}$) and the tidal deformability parameters ($\Lambda_{1}$, $\Lambda_{2}$)~\cite{Hinderer2008} of their components. The presence of matter is predominantly captured by the effective tidal deformability parameter ($\widetilde{\Lambda}$) which is defined by a suitable combination of $m_{1}$, $m_{2}$, $\Lambda_{1}$ and $\Lambda_{2}$~\cite{WadeEtal2014}. Black holes in general relativity are predicted to have zero tidal deformability, i.e., $\Lambda_{1} = \Lambda_{2} = 0$. For a BBH system, this leads to $\widetilde{\Lambda} = 0$ irrespective of their component masses and spins. Moreover, precise estimation of the tidal deformability parameters can constrain the theoretically proposed equations of state of neutron stars and, thus, shed light on the nature of their internal composition~\cite{GW170817-discovery, FlanaganHinderer2008, ReadEtal2009}. 

\begin{table*}[htb]
{\footnotesize %%% fontsize
	\label{table:eos_pe}
	\begin{center}
		\begin{tabular}{ | c | c | c | c | c | c | c | c | c | c | }
			\hline
			\multicolumn{3}{| c |}{\textbf{Source parameters}} &  & \multicolumn{3}{ c |}{\textbf{Measurement accuracies}} & \multicolumn{3}{ c |}{\textbf{Improvements in \%}} \\
			%--------------- 
			\cline{1-3}\cline{5-7}\cline{8-10}
			%--------------- paraameters
			          &                                                   &         &         &                &               &                             &                &               &                             \\
			$m_1,m_2$ & $\Lambda_{1},\Lambda_{2}$ $[\widetilde{\Lambda}]$ & $D_{L}$ & Network & $\Delta M_{c}$ & $\Delta \eta$ & $\Delta\widetilde{\Lambda}$ & $\Delta M_{c}$ & $\Delta \eta$ & $\Delta\widetilde{\Lambda}$ \\
			($\Msun$) &                                                   & (Mpc)   &         & ($\Msun$)      &               &                             & (in \%)        &  (in \%)      &  (in \%)                    \\
			%-----------------------
			\hline
			\hline
			1.35,1.35 & 400,400 [400]   & 40  & LH     & 9.7e-5 & 6.8e-3 & 131.9 & 23.7 & 14.7 & 24.5 \\
			&         &                 & AHL & 7.4e-5 & 5.8e-3 & 99.6   &       &      &    \\
			\hline	                                       
			1.35,1.35 & 857,857 [857]   & 40  & LH     & 9.8e-5 & 6.8e-3 & 156.9 & 15.3 & 14.7 & 26.6 \\
			&         &                 & AHL & 8.3e-5 & 5.8e-3 & 115.2  &       &      &    \\
			\hline	                                       
			1.60,1.17 & 120,980 [551.5] & 40  & LH     & 1.3e-4 & 9.6e-3 & 147.9 & 15.4 & 13.5 & 25.5 \\
			&         &                 & AHL & 1.1e-4 & 8.3e-3 & 110.9  &       &      &    \\
			\hline	                                       
			\hline	                                       
			1.35,1.35 & 400,400 [400]   & 100 & LH     & 1.6e-4 & 7.6e-3 & 269.0 & 18.8 & 16.3 & 19.1 \\
			&         &                 & AHL & 1.3e-4 & 5.6e-3 & 217.7  &       &      &    \\
			\hline	                                       
			1.35,1.35 & 857,857 [857]   & 100 & LH     & 1.7e-4 & 7.5e-3 & 418.7 & 17.6 & 10.6 & 47.8 \\
			&         &                 & AHL & 1.4e-4 & 6.7e-3 & 218.6  &       &      &    \\
			\hline	                                        
			1.60,1.17 & 120,980 [551.5] & 100 & LH     & 1.9e-4 & 1.0e-2 & 384.2 & 15.8 & 10.0 & 44.1 \\
			&         &                 & AHL & 1.6e-4 & 9.0e-3 & 215.0  &       &      &    \\
			\hline	                                       
			\hline	                                       
			1.35,1.35 & 400,400 [400]   & 250 & LH     & 4.7e-4 & 8.5e-3 & 1141.5& 25.5 & 10.6 & 40.8 \\
			&         &                 & AHL & 3.5e-4 & 7.6e-3 & 675.4  &       &      &    \\
			\hline	                                       
			1.35,1.35 & 857,857 [857]   & 250 & LH     & 3.5e-4 & 6.9e-3 & 1298.6& 20.0 & 17.4 & 54.7 \\
			&         &                 & AHL & 2.8e-4 & 5.7e-3 & 588.5  &       &      &    \\
			\hline	                                        
			1.60,1.17 & 120,980 [551.5] & 250 & LH     & 3.2e-4 & 1.0e-2 & 2264.7& 9.4  & 10.0 & 69.9 \\
			&         &                 & AHL & 2.9e-4 & 9.0e-3 & 638.51 &       &      &    \\
			\hline
		\end{tabular}
	\end{center}
	%\vspace{-0.5cm}
	\caption{This table summarizes comparisons of key properties of binary neutron star mergers events with high and relatively low-SNR events. Each of the BNS events is simulated with two neutron star EOSs, one softer (SLy4) and one stiffer (BHB$\Lambda\phi$). The uncertainties in mass parameters, namely in chirp-mass ($\Delta M_{c}$) and symmetric mass-ratio ($\Delta\eta$) as well as uncertainties in the effective tidal deformability parameter ($\Delta\widetilde{\Lambda}$) of the BNS systems are quoted with $90\%$ credible level (see subsection~\ref{sec:eos_pe}). The percentage improvements are quntified following the expresion in Table~\ref{tab:pe} caption.}
} %%% fontsize
\end{table*}

Here we illustrate how the addition of the LIGO-Aundha detector can potentially impact our ability to constrain the effective tidal deformability parameter ($\widetilde{\Lambda}$) as well as discriminate it from the $\widetilde{\Lambda} = 0$ case corresponding to BBHs. We do so by employing a fully Bayesian statistical framework~\cite{GW170817-PE}. 

We analyze a set of simulated BNS events with source properties consistent with the first BNS event, GW170817~\cite{GW170817-discovery, GW170817-PE}. Although the chirp-mass ($M_{c}$) was very well determined to be 1.188 $\Msun$, the component masses have broader uncertainties due to the less precisely measured mass-ratio parameter (e.g., the symmetric mass-ratio, $\eta$). Moreover, although GW170817 could successfully rule out the stiffest equations of state, it still has sufficiently broader uncertainty in estimated tidal deformability parameters, thereby, leaving a wide variety of neutron star EOSs viable.

We perform a systematic injection study of Bayesian parameter estimation for a set of simulated signals from BNS events covering the extreme corners of the parameter space, comprising component masses and tidal deformability parameters that are consistent with GW170817. Initially, we consider all the sources to be located at a luminosity distance ($D_{L}$) of 40 Mpc (similar to the GW170817 event). We consider one equal-mass BNS ($m_1 = m_2 = 1.35\Msun$) and one unequal-mass BNS ($m_1 = 1.60\Msun, m_2 = 1.17\Msun$), with soft EOS, namely SLy4~\cite{RefSLy4}, consistent with the GW170817 observation. For the equal-mass case, we also consider the possibility that the neutron stars have a stiff EOS, namely, BHB$\Lambda\phi$~\cite{RefBHBLp}. 

Given the current estimation of BNS merger event rates (see Sec.~\ref{sec:rates}), it is improbable that such an event will be observed at a distance $D_{L} \lesssim 40$ Mpc in the near future. We, therefore, perform additional simulations 
%for Bayesian parameter estimation considering the 
with the entire set of events (a) at $D_{L} = 100$ Mpc as well as (b) at $D_{L} = 250$ Mpc. In our simulations, we use \textsc{IMRPhenomDNRTidal} waveform model~\cite{DietrichEtAl-TidalWF} for the coalescing BNS systems with slow ($|s_{1}|, |s_{2}| \leq 0.05$), in-plane component spinning configurations for simplicity since astronomical distributions demonstrate that more rapidly spinning BNS systems are rare as well as this configuration captures the key aspects reasonably well. We summarize the measurements of mass and tidal deformability parameters with 90\% Bayesian uncertainty intervals in table~\ref{sec:eos_pe}. 

This study demonstrates that for the very high SNR events (with comparable single detector SNRs $\sim 110 - 130$ in each of the H, L, and A detectors) the improvement of precision in $M_{c}$ is in the range of $15 - 25\%$ and in $\eta$ is of about $10 -17\%$ for the AHL-network of detectors as compared to the HL-network. The improvement in $\widetilde{\Lambda}$ estimation in favor of the AHL-network relative to two US-based detectors is also nominal -- at about $25\%$. As the source distance increases resulting in a decrease in SNR, the improvement in precision for $D_{L}$ and $\eta$ does not change much for comparable SNRs in the three detectors. However, we find that for low SNR events the precision in $\widetilde{\Lambda}$ improves significantly. For the set of BNS observation at $D_{L} = 100$ Mpc, we find that improvements can be in the range of $20\%$ to $45\%$. For the more distant sources, e.g., at $D_{L} = 250$ the improvements are generally more than $40\%$, and can be as high as $70\%$ in favor of AHL relative to the HL-network. Moreover, for such distant sources, the lack of precision in  $\widetilde{\Lambda}$ can render it difficult to rule out the BBH-case corresponding to $\widetilde{\Lambda} = 0$, particularly for the soft (SLy4-like) EOS. (As a comparison to the range of $\widetilde{\Lambda}$ parameter for different theoretically motivated neutron star EOS models please refer to~\cite{EOSModelSelectionLVC}.) Thus, for the events with relatively weak signals -- which will be at farther distances and, hence, in relatively abundant numbers -- the source classification (i.e., BBH {\it vs} BNS/NSBH) will get significantly enhanced. This will be important for generating alerts for the subsequent follow-up with astronomical observations across the electromagnetic spectrum.

%==============================================================================================
\section{Sky localization and early warning}
\label{sec:sky_localization}

\def\patch{\textit{patch}}
\def\sqd{\textit{sq.deg.}}
\def\Msn{M_\odot}
%\cmt{Assigned to: Javed \& co}

One of the main advantages of expanding the HL network to include LIGO-Aundha is
that it substantially improves the localization of CBCs in the
sky~\cite{Ligo-india-Fairhurst-2014, Pankow:2019oxl}. BNSs and a fraction of NSBHs have long
been expected to produce prompt counterparts and afterglows in all
electromagnetic (EM) bands. For BNS mergers in particular, it has been
hypothesized that the post-merger central engine can launch short gamma-ray
bursts (sGRBs)~\cite{Lattimer:1976kbf, Lee:2007js}, kilonovae~\cite{Li:1998bw,
Metzger:2010sy}, and radio waves and X-rays before and after
merger~\cite{Nakar:2011cw, metzger2012most, Metzger:2016mqu, Hallinan:2017}.

%They also give a very neutron-rich environment, outflowing towards the interstellar medium (cite).
These emissions carry information about both the progenitors -- e.g., the
equation of state of neutron stars -- and the circum-merger environment. The
prompt, and often, transient emission on the one hand and the late-time
afterglow on the other hand complement each other in conveying that
information, as was demonstrated amply by the multi-messenger observations of
the binary neutron star event, GW170817~\cite{GBM:2017lvd}. The joint
observation of GWs followed by the sGRB,  GRB~170817A, and the kilonova
AT~2017gfo,~\cite{GBM:2017lvd} confirmed the several-decade-old hypothesis
that compact object mergers were progenitors of these exotic transients.
However, GW170817 is so far the only gravitational-wave event to be observed in
other channels. Improvements in GW detectors and expansion of the GW network is
therefore required to realize more multi-messenger observations and expand our
knowledge about the physical processes that occur in these systems.

The chances of telescopes spotting that EM emission improve if the localization
area in the sky associated with the GW signal is small. This is particularly true for
tracking down optical counterparts since the fields of view (FOV), or
beam sizes, of these telescopes are small (sub-arcminutes) compared to the the
typical GW sky-localization area. The small localization with the rapid search
strategies~\cite{Rana:2017a, Coughlin:2018e, Gosh:2016e} can enhance the probability
of finding the optical counterpart of the GW source. For prompt and transient emission,
a narrow sky-area implies a small number of telescope slews and a quicker locking
on to the target before it fades ~\cite{Rana:2019a}. 
%In the case of 
The search for kilonovae and prolonged afterglows is aided by narrow sky-areas 
since they are scannable quickly by telescopes and make multiple observations
of the same telescope fields of view in those areas more feasible. This, in turn,
improves the probablity for spotting their onset.
In the case of larger localizations, the early observations are likely to be missed. 
In some cases (e.g., GW190425~\cite{Abbott_2020}), large localizations can prohibit identification of the EM counterpart 
entirely. 
% observe can help since they allow observation of the EM-candidate over multiple epochs. 
Also, radio follow-up affords complementary
observations for day-time and dust-obscured events, where the hunt in optical is
difficult. In that case, a small volume in 3D localization is important for
the galaxy targeted radio observation to get arcsecond localization~\cite{Rana:2019b}.
% to remove the astrophysical false positives and increase the detections confidence.

Moreover, if the sky-localization is sharper, then spectroscopy becomes
possible, which can provide not only clues on the progenitor composition but
also the redshift of the event. Spectroscopy requires longer exposure times. 
A narrow sky error region implies a smaller number of fields of view to search in 
for finding the counterpart. This allows for a quicker homing in on potential 
counterparts and, therefore, extended exposures thereafter.
The first discovery of the optical-counterpart
of the event GW170817 was after $\sim$11 hours of the GW trigger. Detection of the
EM-counterparts must be much quicker if their prompt emissions are the desired
target.

To demonstrate the benefit of including LIGO-Aundha in the GW network, we
simulate a population of binary neutron stars and compare the distribution of
GW localization in the two detector networks: HL and AHL. We generate a
population of 9,308,544 simulated BNS signals using the
\texttt{TaylorF2}~\cite{sanjeev1991, BDIWW95, Blanchet:2005tk,
PN-Bounanno-comparison} waveform model. Both source-frame component masses are
drawn from a Gaussian distribution between $1.0 \, M_\odot$ $< m_1, m_2 <$ $2.0
\, M_\odot$ with mean mass of $1.33 \, M_\odot$ and standard deviation of $0.09
\, M_\odot$, modeled after observations of galactic BNSs~\cite{Ozel:2016oaf}
(note, however~\cite{Abbott_2020}). The component spins are aligned or
anti-aligned with respect to the orbital angular momentum with the
dimensionless spin amplitude on the neutron stars restricted to $0.05$,
motivated by the low spins of BNSs expected to merge within a Hubble
time~\cite{Burgay:2003jj,Zhu:2017znf}. The signals are distributed uniformly
in sky, orientation, and comoving volume up to a redshift of $z =0.4$.
%We tested the performance of two, three and five ground-based GW detectors by generating localizations for BNS merger events. The two networks we have considered are HL and AHL. We assumed the detectors are running with there design sensitivity. In our simulation, we did injections of 1260 BNS sources of 1.4 and 1.4 $\Msn$ at distance from 30 Mpc to 300 Mpc. If $\iota$ is the orbital inclination of the source, then the injections are uniform over $cos(\iota)$. The injections are distributed isotropically on the sky. For simplicity, we kept all the sources are non-spinning neutron stars.
We simulate the GW signal and calculate the expected SNR in Gaussian noise
considering the three LIGO detectors at A+ sensitivity for each BNS. We mimic
the results from a matched-filter GW search pipeline (current low-latency
matched-filter searches running on LIGO-Virgo data include
\texttt{GstLAL}~\cite{Sachdev:2019vvd,Hanna:2019ezx},
\texttt{PyCBCLive}~\cite{Nitz:2018rgo},
\texttt{MBTAOnline}~\cite{Adams:2015ulm}, \texttt{SPIIR}~\cite{chu2017low}) by
considering the signals that pass a network SNR threshold of 12.0
%(\gv{BBH rate uses 8})
to be `detected'. We then calculate the sky-localization posteriors for the detected candidates
using a rapid Bayesian localization tool,
\texttt{BAYESTAR}~\cite{PhysRevD.93.024013}. We use the most
recent BNS local merger rate from~\cite{Abbott_2020} of $320^{+410}_{-240}
\, \rm{Gpc}^{-3}\rm{yr}^{-1}$ to estimate the number of events detected per
year in the detector network.
% Repeated from above: In this section, we discuss the improvement of the localization after adding LIGO-India in the network with the other GW-detectors. The localization from two LIGO detectors are very curse, not depends much on the location of the source in the sky.

%When comparing the simulated BNSs recovered with HL and AHL networks, we first find a noticeable improvement in their detection rates.
%First we discuss the improvement in the rate of recovery of the BNS mergers. Once the recovery of 1260 injected BNS sources is done by three different networks, we have the estimate of the rate of recovery.
%Figure~\ref{fig-recobar} shows the percentages of recovered sources by those networks. Out of a total 1260 sources, HL recovered 915 sources (or $75.2\%$) and AHL recovered 1184 (or $94.7\%$). It is precisely visible that the rate of recovery gives  $\sim20\%$ jump from HL to AHL network. However, the AHL network could not recover all the sources because the inclinations of some of them are such that the sources do not register enough SNR to be detected by it.

%\tcr{Note:Shoud we show the sensitivity of AHL network at deifferent location on the sky according to the localization size?}

\begin{figure}
	\begin{subfigure}[t]{0.49\textwidth}
	\includegraphics[width=1.0\textwidth]{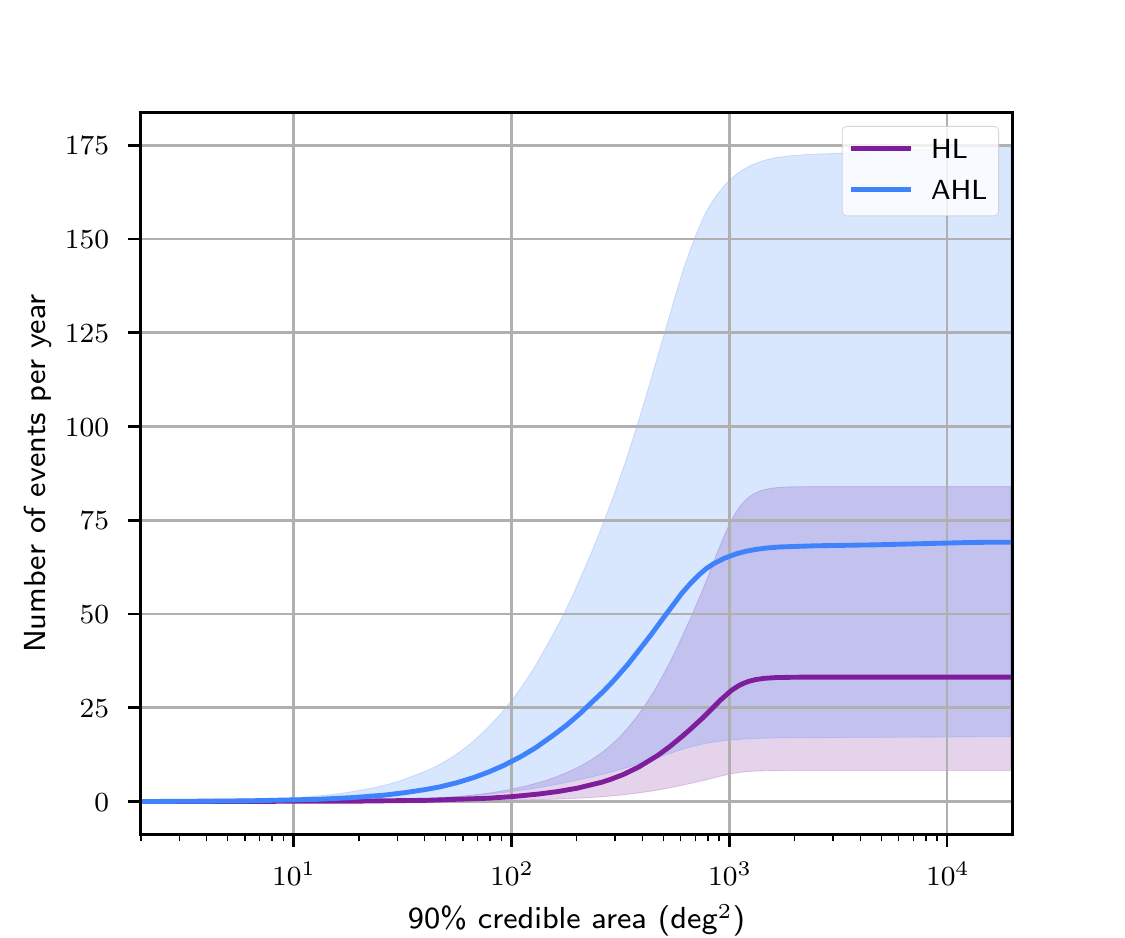}
	\end{subfigure}%
	\begin{subfigure}[t]{0.49\textwidth}
        \includegraphics[width=1.0\textwidth]{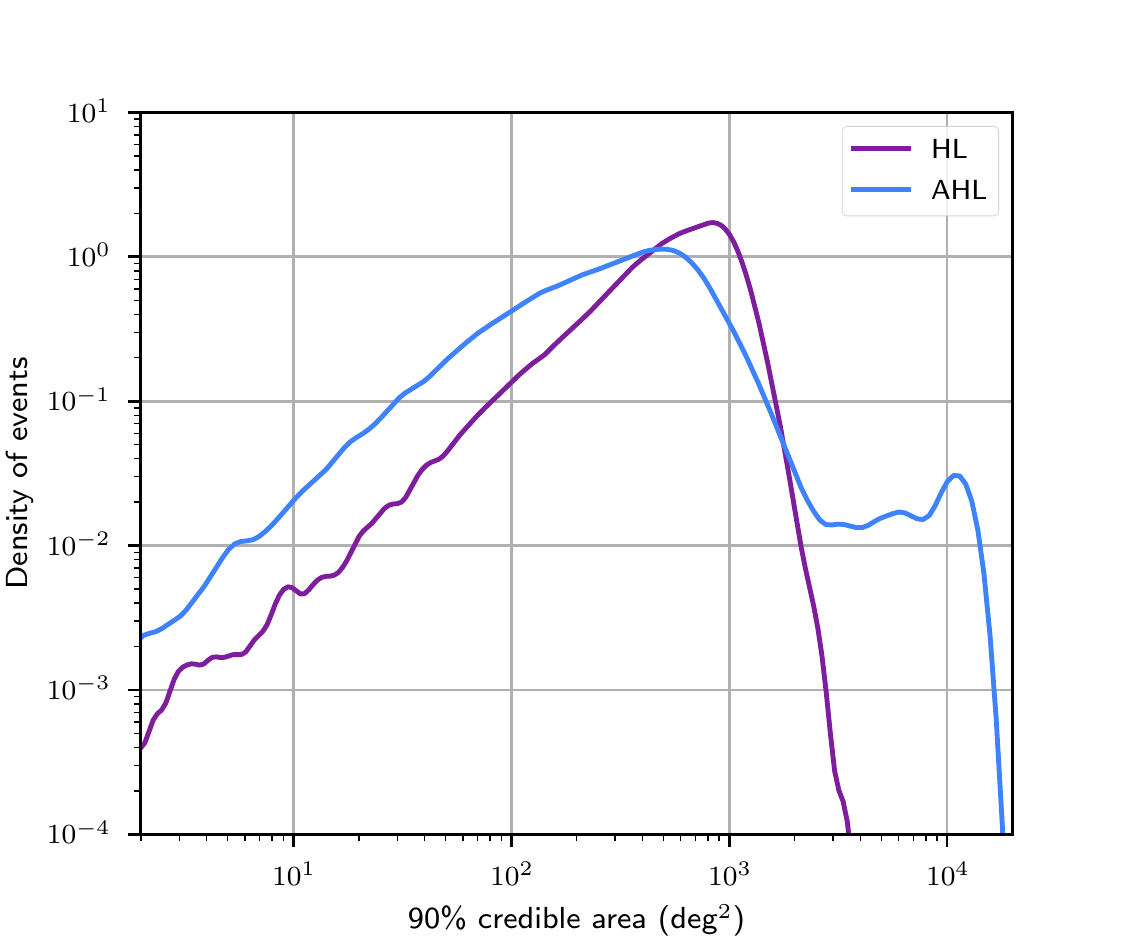}
        \end{subfigure}%
	\caption{Distributions (left: cumulative,
right: density) of the sky localizations (90\% credible interval) of the BNSs
that pass the fiducial SNR threshold of 12 for the two detector networks: HL
(purple) and AHL (blue). Using the latest median BNS merger rate from~\cite{Abbott_2020} of $320\, \rm{Gpc}^{-3}\rm{yr}^{-1}$, we find that the HL (AHL) network is expected to detect $\sim$ 33 (69) events per year. The shaded regions represent the uncertainty in the BNS merger rate estimate.}
	\label{fig-hlvki90}
\end{figure}

In \Cref{fig-hlvki90}, we show the distributions (left: cumulative,
right: density) of the sky localizations (90\% credible interval) of the BNSs
that pass the fiducial SNR threshold of 12 for the two detector networks: HL in
purple and AHL in blue. The shaded regions show the uncertainty in the number of
detections due to the uncertainty in the current local BNS merger rate of
$320\, \rm{Gpc}^{-3}\rm{yr}^{-1}$~\cite{Abbott_2020}. The improvement due to the
addition of LIGO-Aundha to the network is clearly visible in this figure; the
AHL network detects about twice (17 -- 175) as many signals as the HL (8 -- 84) network. Further the
peak of distribution for HL is around $800 \,\rm{deg}^2$, about twice that for
the AHL network.

%Figure~\ref{fig-areaarea-dist} shows the performances of all three networks. As expected, AHL performs much better than HL to localize an event. In this scatter plot every scattered point represents a GW event. Both the axes represent the area of a \patch\ for two different networks. Plot in the top panel shows the comparison between HL and AHL.

\begin{figure*}
	\centering
	\includegraphics[width=\columnwidth]{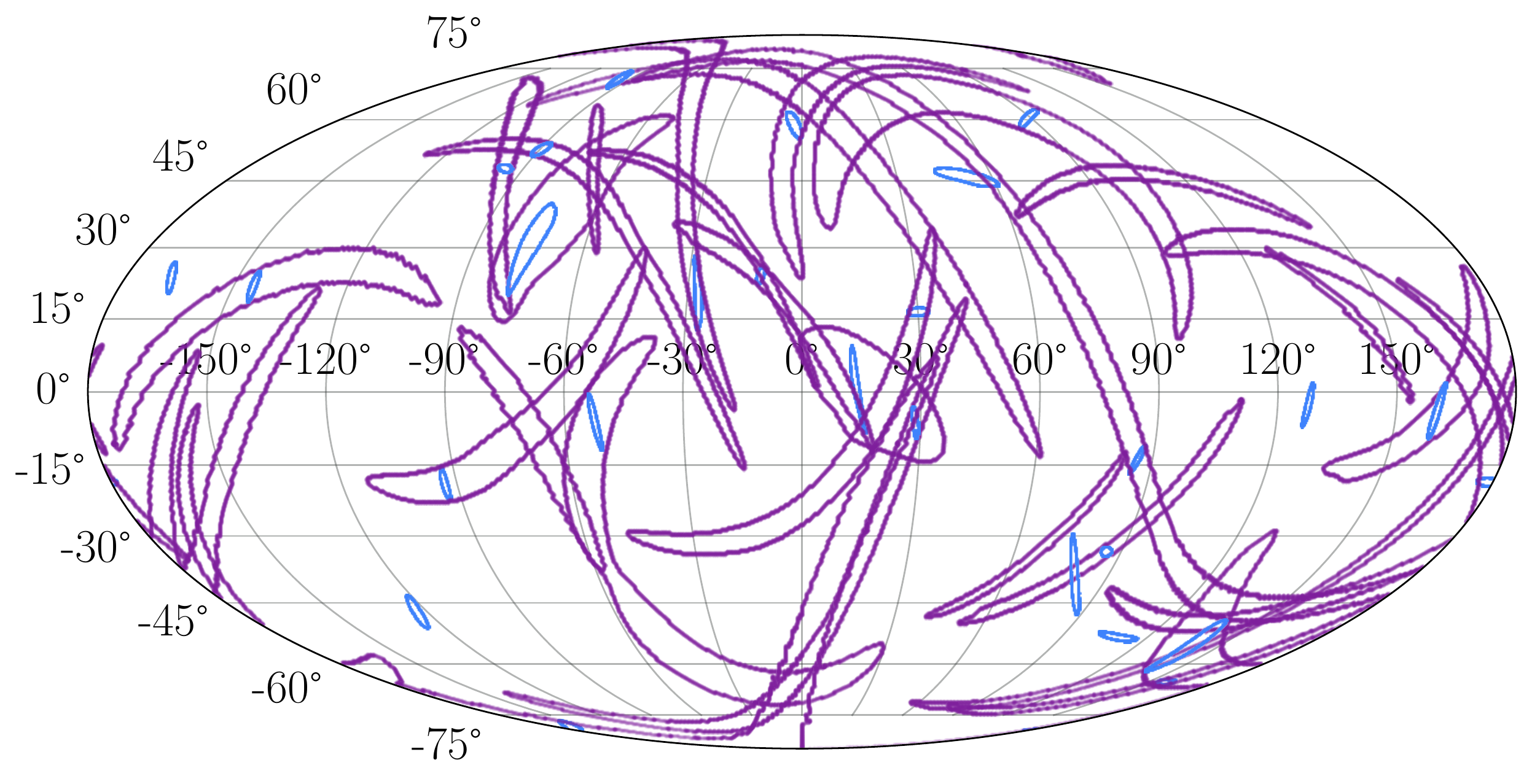}
	\caption{Shown above are the $90\%$ probability contours of source localization on the sky.
          The blue and purple contours correspond to the networks AHL and HL, respectively, for the same
          sources selected from our BNS simulations, and exhibit the improvement achieved due to the 
          inclusion of the third detector in LGN.
          Note that AHL detects every signal that is detected by HL.
          The isolated blue contours correspond to sources detected by AHL but not by HL.}
	\label{fig-skyplot}
\end{figure*}

\Cref{fig-skyplot} shows the shape and the areal projection of several localizations on the sky. The HL (AHL) localizations are shown in purple (blue) contours. 
%They are long arc-shaped and larger in area.
%The red and blue contours are for AHL and HLVKI networks respectively.
Most of the HL localizations are long arcs that extend over both hemispheres. When we include LIGO-Aundha, the degeneracy breaks and the localizations typically shrink to one of the hemispheres. In this plot, a few blue contours have no corresponding purple contours. These are the events detected by AHL
%and HLVKI
but not by HL. The orientations of the HL localizations are concentric. On the other hand AHL localizations are randomly oriented. This is because a single baseline offers a single time delay for any event, which in turn is consistent with source sky-positions that all lie on a single circle in the sky. Of course, detector antenna functions help localize the source position further in those circles, thereby, reducing the localization area to arcs. The addition of a third detector provides additional time delays that aid in reducing the error patches further, as evident in the blue error contours.

%Figure~\ref{fig-sensitivity-map} is a full sky map of the localization sensitivity for AHL network. The map shows the relative size of the area of the \patch\ around the different region on the sky. To generate this map we injected BNS sources of 1.4 and 1.4 $\Msn$ on a uniform grid on the sky at 200 Mpc distance. We have chosen a grid of $10^{\circ}$ gap along longitude and $5^{\circ}$ gap along latitude. The color code is normalized, shown in the colorbar. In the map, the sky above Europe has darker red color than the sky above USA. This implies that the area of localization from the same event will have $\sim 5$ times larger area on Europe sky than USA sky. The two darkest regions on the two hemispheres will have the largest localization for AHL network.

\begin{figure*}
	\centering
	\includegraphics[width=\columnwidth]{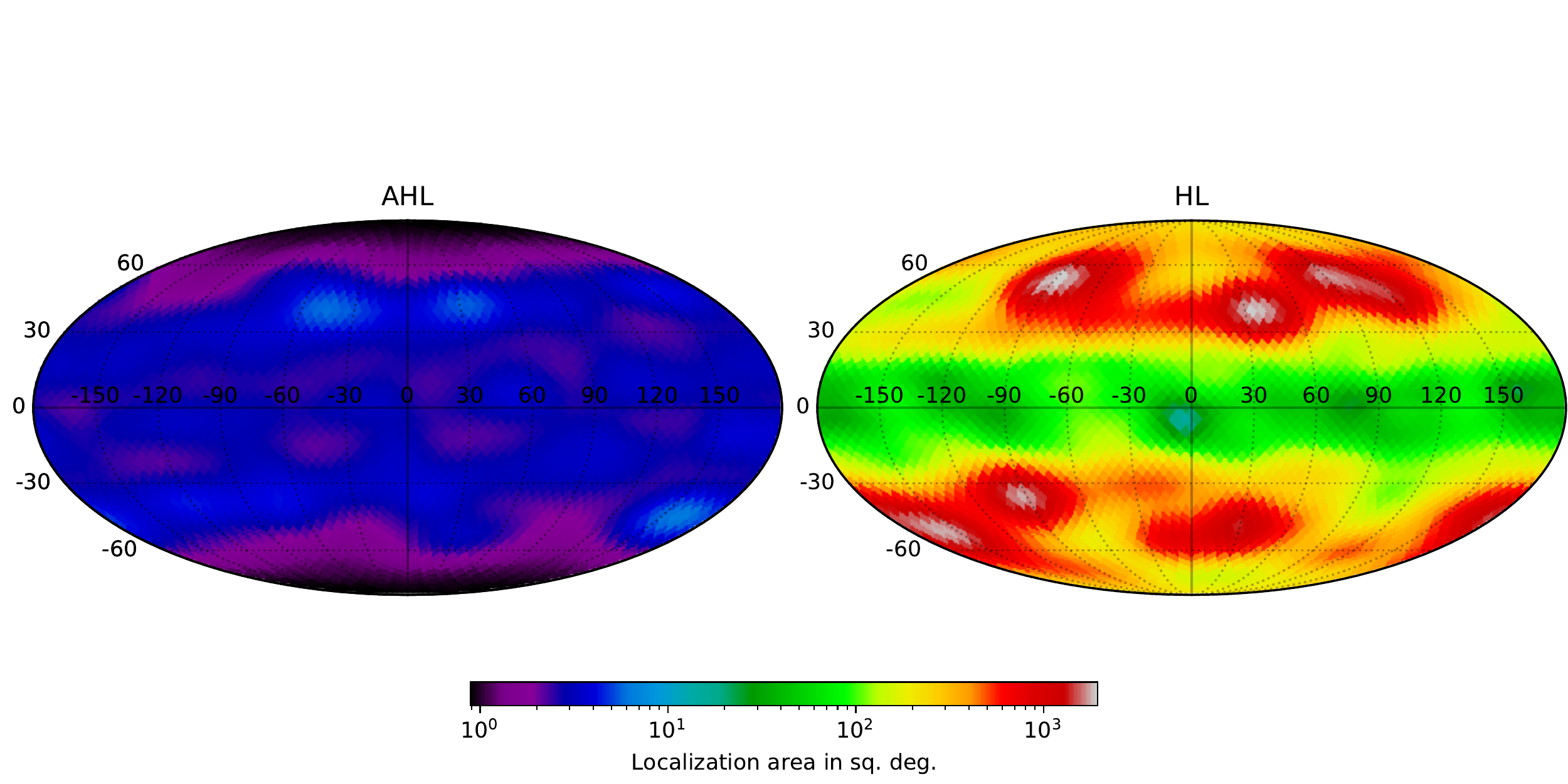}
	\caption{Localization area for BNS sources across the sky for the AHL and HL networks. 
        Typically, smaller areas for AHL compared to HL show the degree of improvement the presence 
        of the third LIGO detector in the LGN brings about. 
          The sky map is divided into equal-area pixels in the HEALPix format. Each pixel has one injected BNS source in it and the colorbar value of the pixel is the $90\%$ probable area of the localization of the source calculated by \texttt{BAYESTAR}.
          All the BNSs are injected at 100 Mpc, and have $m1=m2=1.4 \Msn$ and $5^{\circ}$ inclination. 
        }
	\label{fig-sensitivity-map}
\end{figure*}

The study in \Cref{fig-sensitivity-map} compares gravitational-wave sky-localization by HL and AHL for similar sources in different parts of the sky. For this purpose, we divide the entire sky into equal-area pixels in HEALPix~\footnote{A HEALPix map parameterized by the variable NSIDE is a representation of the full-sky filled with $N_{pix}=12\times NSIDE^2$ equal-area pixels. The value corresponding to a pixel is the probability of finding the GW source in that pixel in the sky.} format of NSIDE 16~\cite{healpix:2005}. We inject one BNS source in each of the 3072 pixels and calculate the localization area of the source using \texttt{BAYESTAR} for HL and AHL. All the injected sources are of $m_1,m_2=1.4 \Msn$ at a distance of 100 Mpc and have an orbital inclination of 5 deg. In \Cref{fig-sensitivity-map}, the colorbar represents the area of the $90\%$ credible region of the localization in $\rm{deg}^2$. In other words, the pixel value is the $90\%$ probable area of the localization of the source injected in that pixel. For the HL network, the smallest localization area is $\sim 45 \,\rm{deg}^2$, and the largest one is $\sim 1732\,\rm{deg}^2$. On the other hand, the smallest and the largest localization areas for AHL are $\sim 1\,\rm{deg}^2$ and $\sim 21\,\rm{deg}^2$. 
Compared to HL, the AHL sky areas are not smaller by a constant factor. How much the area shrinks depends on the true position of the source.  The results in \Cref{fig-skyplot} and \Cref{fig-sensitivity-map} are consistent and are two different representations of sky-localization analysis using the same aforementioned code.
%REPEATED FROM ABOVE: We compare AHL and HL by the number of BNS sources they localize within a given credible localization area. %To do that, we take the same sources and pixelized position on the sky as in the study of \Cref{fig-sensitivity-map}. We inject %3072 sources in the pixelized sky of NSIDE of 16 at a distance of 100 Mpc. 
Furthermore, in \Cref{fig-sensitivity-map} we observe that the AHL network localizes all the 3072 events with $90\%$ credible area within 20 sq.deg. On the other hand, the HL localizes 793 events  within 100 $\rm{deg}^2$ and 239 events within 50 $\rm{deg}^2$. 
%TO BE SAFE LEAVE IT OUT FOR NOW: In \Cref{event-coverage} we show the changes of the number ratio of localized events with localization areas from 20 $\rm{deg}^2$ to 100 $\rm{deg}^2$. The red dots represent the ratio $N_{AHL}/N_{HL}$ within the specific area, where $N_{AHL}$ and $N_{HL}$ are the numbers of events localized within a given credible localization area respectively by AHL and HL. 

%\begin{figure*}
%	\centering
%	\includegraphics[scale=.5]{recover-bar.pdf}
%	\caption{The bar plot shows the fraction of nonspinning binary neutron stars, with component masses 1.4 and 1.4 $\Msn$, recovered by the HL and AHL networks. The number of signals injected is 1260, out of which 866 events (up to the right-edge of the green bar) are recovered by HL and 1091 events (up to the right edge of the red bar) are recovered by AHL.}
%	\label{fig-recobar}
%\end{figure*}

%\begin{figure}
%	\centering
%	\includegraphics[scale=.2]{triangular-alllog-m.pdf}
%	\caption{Nonspinning BNS with 1.4 and 1.4 $\Msn$. Each scattered point corresponds the $90\%$ area of the localization for an GW event. x-axis and y-axis represent the $90\%$ area from two different networks. The three plots are the comparison of area vs area for all events for  three networks. The colorbar show the distance distribution of the sources. {\color{red} [Please remove or replace figure.]}}
%	\label{fig-areaarea-dist}
%\end{figure}

%\begin{figure*}
%	\centering
%	\includegraphics[width=\columnwidth]{allsky_coverage_ratio.pdf}
%	\caption{Exhibits the improvement achieved by AHL for the large fraction of events to localize within smaller areas than %HL. Within 100 $\rm{deg}^2$, HL localizes  793 sources, and AHL localizes 3072 sources. Likewise, 239 sources are localized HL %and 3072 are localized by AHL within 50 $\rm{deg}^2$. All the BNSs are injected in pexelized sky of NSIDE 16 at 100 Mpc, and %have $m1=m2=1.4 \Msn$ and $5^{\circ}$ inclination.}
%	\label{event-coverage}
%\end{figure*}

%\begin{figure}[h]
%    \hspace{1.5cm}
%    \centering
%    \includegraphics[scale=0.4]{figures/ahl_locmap.pdf}
%    \includegraphics[scale=0.4]{figures/hl_locmap.pdf}
%    \caption{Localization area for BNS sources across the sky for the AHL and HL networks. Typically smaller areas for AHL compared to HL show the degree of improvement the presence of the third LIGO detector in the LGN brings about. The sky map is divide into equal-area pixels in HEALPix format of NSIDE 16. Each pixel has one injected BNS source in it and the value of the pixel, as can be read off from the color-bar, is the $90\%$ probable area of the localization of the source calculated by BAYESTAR. All the BNSs are injected at 50 Mpc, and have $m1=m2=1.4 \Msn$ and $5^{\circ}$ inclination. The darker red pixels represent larger in localization area.}
%    \label{fig-sensitivity-map}
%\end{figure}

\subsection{Early warning of binary neutron star mergers}
August 17, 2017 saw the beginning of a new era in multi-messenger astronomy.
The joint detection of GWs by the LIGO and Virgo interferometers and the sGRB
by the Fermi-GBM and INTEGRAL satellite from the BNS coalescence,
GW170817~\cite{GW170817-discovery, GBM:2017lvd} confirmed the long-standing hypothesis
that compact object mergers were progenitors of short GRBs. Apart from the
gamma-ray burst, which was observed $\sim 2$~s after the merger event, the
first manual follow-up observations took place $\sim 8$ hours after the epoch
of merger~\cite{GBM:2017lvd}. This delay was caused by the delay in sending out
GW information: the GW alert was sent out $\sim 40$ minutes ~\cite{GCN21505},
and the sky localization $\sim 4.5$~hours~\cite{GCN21513} after the signal
arrived on earth. By the time EM telescopes participating in the follow-up
program received the alerts the source was below the horizon for them.

For a fraction of BNS events it will be possible to issue alerts up to $\delta
t \sim 60\,\rm s$ before the epoch of merger~\cite{Cannon:2011vi,
10.1093/mnras/stw576, Akcay:2018aqh}. Pre-merger or \textit{early warning}
detections will facilitate electromagnetic observations of the prompt emission,
which encodes the initial conditions of the outflow and the state of the merger
remnant. Early optical and ultraviolet observations will be key to our
understanding of \textit{r}-process nucleosynthesis~\cite{Nicholl:2017ahq} and
shock-heated ejecta~\cite{Metzger:2017wot}, while prompt X-ray emission would
reveal the final state of the
remnant~\cite{Metzger:2013cha,Ciolfi:2014yla,Siegel:2015twa}.  Early
observations made in the radio band could indicate pre-merger magnetosphere
interactions~\cite{Most:2020ami}, and might be able to test models that
predict BNS mergers as a possible precursor of fast radio
bursts~\cite{Totani:2013lia,Wang:2016dgs,Dokuchaev:2017pkt}. Early-warning GW alerts have recently been implemented~\cite{Sachdev_2020, Nitz:2020vym} and also demonstrated~\cite{magee2021demonstration} recently.

Here we will compare the prospects of early-warning detection of GWs from BNSs
for the two detector networks: HL and AHL. We follow the framework laid out
in~\cite{Sachdev_2020}, which implemented an early warning GW pipeline using
the \texttt{GstLAL} matched-filtering software suite~\cite{cannon2020gstlal}.
In particular, we consider 6 different discrete frequency cut-offs
%(\gv{upper frequency cut-off }):
$29\thinspace\rm{Hz}$, $32\thinspace\rm{Hz}$, $38\thinspace\rm{Hz}$,
$49\thinspace\rm{Hz}$, $56\thinspace\rm{Hz}$, and $1024\thinspace\rm{Hz}$ to
analyze signal recovery at (approximately) $58\thinspace\rm{s}$,
$44\thinspace\rm{s}$, $28\thinspace\rm{s}$, $14\thinspace\rm{s}$,
$10\thinspace\rm{s}$, and $0\thinspace\rm{s}$ before merger. We use the
population of BNSs described earlier and the same criteria for `detected'
signal, the signals that pass an SNR threshold of 12.0 in each frequency
configuration are detected with the corresponding pre-merger latency. We then use \texttt{BAYESTAR} to localize all the detected signals for each frequency configuration.

\begin{figure}[h]
        \begin{subfigure}[t]{0.49\textwidth}
        \includegraphics[width=1.0\textwidth]{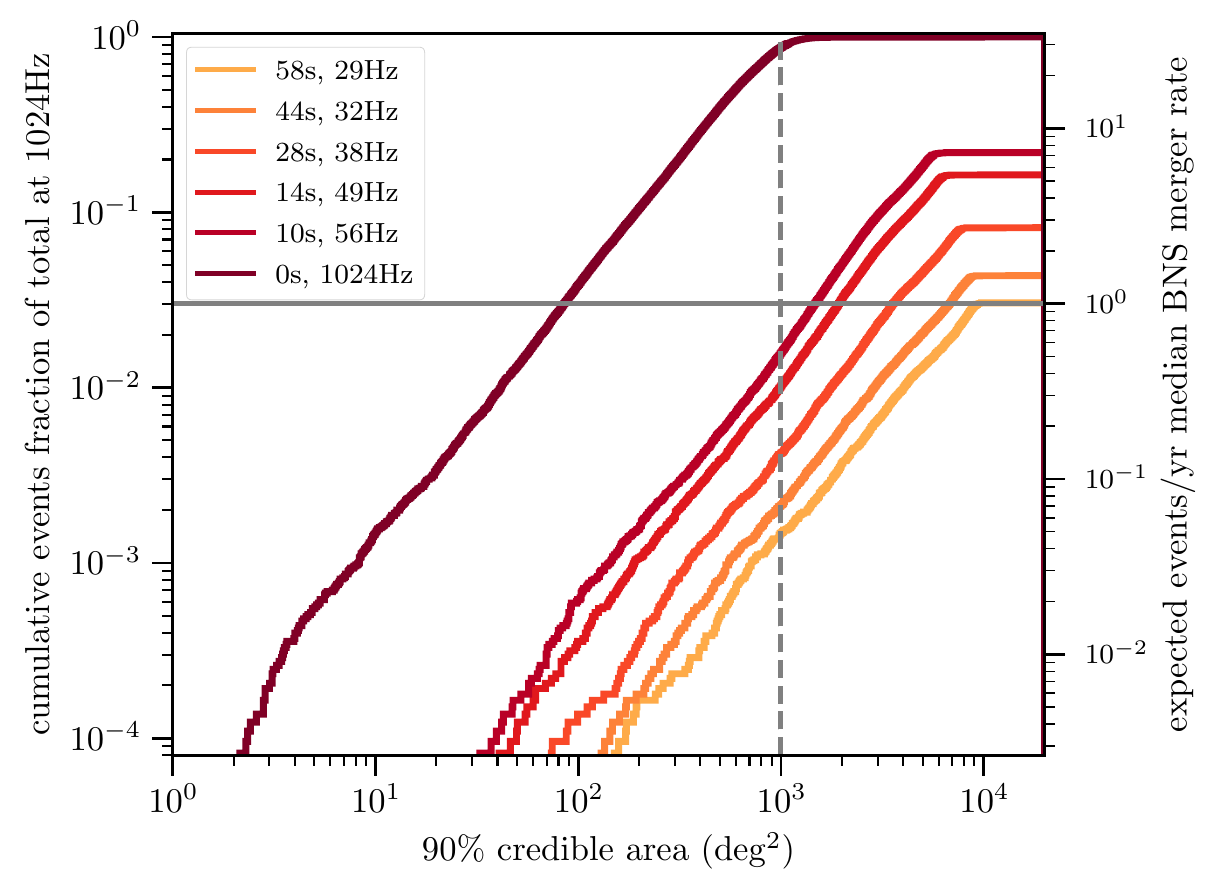}
        \end{subfigure}%
        \begin{subfigure}[t]{0.49\textwidth}
        \includegraphics[width=1.0\textwidth]{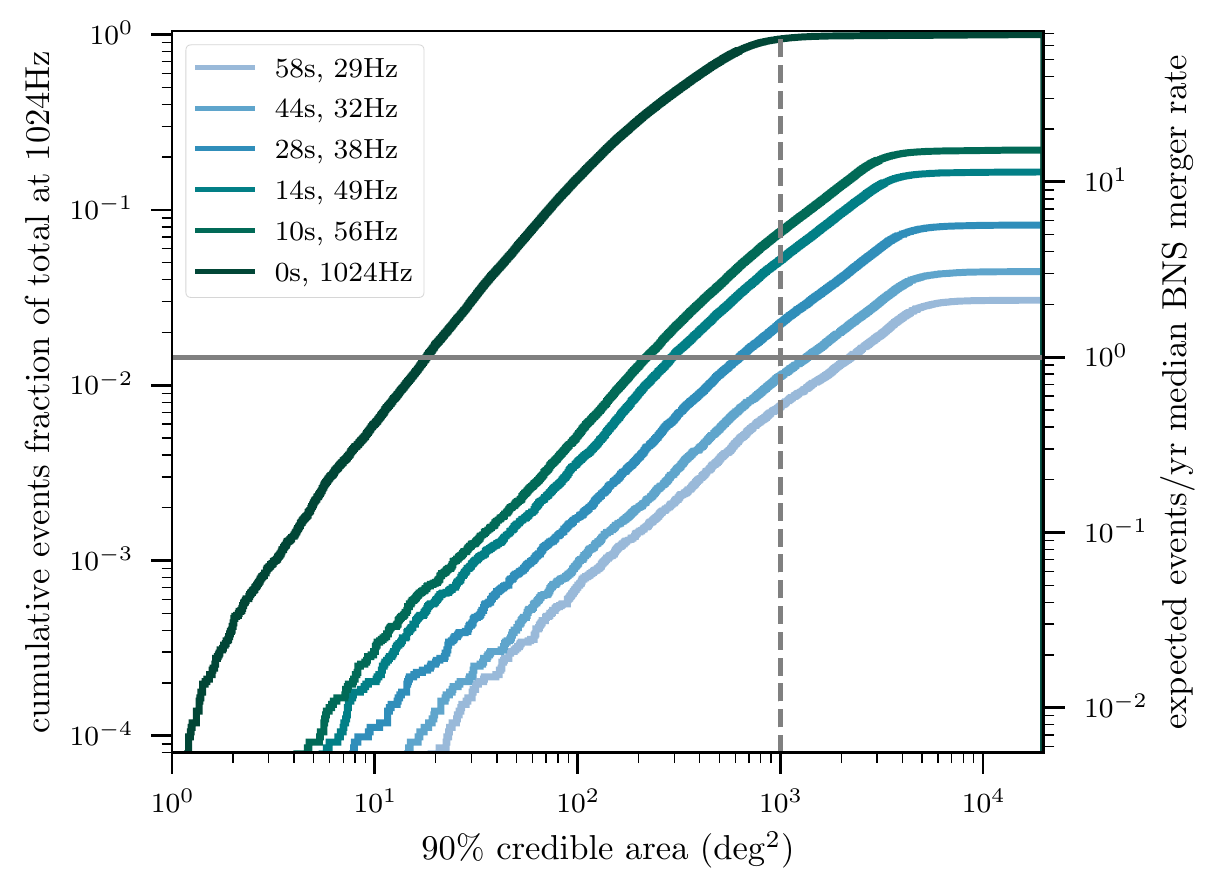}
        \end{subfigure}%
	\caption{Cumulative distributions of the sky localizations (90\%
credible interval) of the BNSs that pass the fiducial SNR threshold of 12 for
the two detector networks: HL (left) and AHL (right). The different colors show
the different frequency bandwidths or pre-merger times considered in our work.
Using the latest median BNS merger rate from~\cite{Abbott_2020} of $320\,
\rm{Gpc}^{-3}\rm{yr}^{-1}$, we find that the HL (AHL) network is expected to
detect $\sim$ 7 (15) mergers per year at least 10 s before merger. Out of
these, the HL network is expected to detect one event every two years before
merger that is also localized to $1000 \,\rm{deg}^2$, while the AHL network is
expected to detect $\sim 5$ events every year before merger that are also
localized to $1000 \,\rm{deg}^2$. }
        \label{fig-ew}
\end{figure}

Our results are shown in \Cref{fig-ew}. For both networks (left: HL, right:
AHL), we show the cumulative distributions of the 90\% credible intervals of
the sky localizations for each pre-merger time considered in our simulation.
The $y$-axis is translated to number of detections per year assuming the current
median BNS merger rate estimate and 100\% duty cycle of the networks. We note that at each
frequency and pre-merger time configuration, the AHL network is able to detect
about twice the number of events as compared to the HL network. In particular,
the number of events per year that could be detected at least 10 s before
merger is 7 (15) for the HL (AHL) network. Adding LIGO-Aundha to the network
will also greatly reduce the area in the sky to which these events can be
localized, thereby vastly improving prospects of observing EM emissions before
and/or at merger. The HL network is expected to detect one event every two years before
merger that is also localized to $1000 \,\rm{deg}^2$, while the AHL network is
expected to detect $\sim 5$ events every year before merger that are also
localized to $1000 \,\rm{deg}^2$.

%\gv{Can we fix secondary y axis limit same}
These localizations are quite large for optical telescopes, which generally
have very small FOVs in comparison. With the current estimates of the BNS merger rate, the AHL network can detect 1 event every $\sim 4$ years 15 s before merger localized to $100 \,\rm{deg}^2$. For the HL network, such an event will be detected once every $\sim 85$ years.

Some of the largest field telesopes, such as the BlackGEM array ($0.65\, \rm
m /2.7 \, \rm{deg}^2$ per telescope) with 3 telescopes planned in the first
phase of operation eventually expanding to 15 telescopes~\cite{blackgem}, the
Zwicky Transient Facility ($1.2 \, \rm m / 47 \, \rm{deg}^2$)~\cite{ztf},
the Dark Energy Camera ($4\, \rm m / 3.8 \,
\rm{deg}^2$)~\cite{Flaugher_2015}, the Rubin Observatory ($8.4 \, \rm m /
9.6 \, \rm{deg}^2$)~\cite{lsst}, the Swope Telescope ($1\, \rm m / 7 \,
\rm{deg}^2$)~\cite{swope}, the Subaru Telescope ($8.2 \, \rm m/1.7 \,
\rm{deg}^2$)~\cite{subaru}, etc. all have FOVs only a small fraction of the GW localizations. Therefore, adding LIGO-Aundha to the LGN will enhance the chances of observing any pre-merger and prompt emissions. Similarly, a larger network involving LGN as well as non-LGN detectors is expected to boost these chances further.

\section{Tests of GR}
\label{sec:tog}
%	\cmt{Assigned to:  Aditya Vijayakumar, P. Ajith, Srashti, and Saleem: The section is ready to read}

% Section is ready for reading.
% NUMBERS CAN BE FOUND IN scripts/TGR_improvement_with_HLA.ipynb
\subsection{Improved constraints on deviations from GR}
%	\cmt{Assigned to:  Aditya Vijayakumar, P. Ajith}
Most of the tests of general relativity performed in \cite{TheLIGOScientific:2016src,LIGOScientific:2019fpa,Abbott:2020jks} are null tests in the sense that they look for deviations around the expectation from GR. These deviation parameters can be measured from each detected event and, assuming certain properties, can be combined across multiple events to get tighter constraints. Measurement of deviations in the post-Newtonian parameters \cite{yunes2009fundamental,TIGER2012,TIGER2012further,Agathos:2013upa}, inspiral-merger-ringdown consistency test \cite{Hughes:2004vw,Ghosh:2015jra,Ghosh:2016xx}, and the upper bound on the mass of the graviton~\cite{Mirshekari:2011yq}, etc., fall under this category.

We illustrate the potential of getting tighter bounds on the GR deviation parameters with the AHL network using the measurement of the mass of the graviton. In GR, GWs are non-dispersive. Hence the corresponding force carrier, graviton, should have zero rest mass.  But there are alternative theories to GR that permit a non-trivial dispersion relation. We assume the following phenomenological form \cite{Yunes:2016jcc,Mirshekari:2011yq,Will:1997bb, Calcagni:2009kc,AmelinoCamelia:2002wr,Horava:2009uw,Sefiedgar:2010we,Kostelecky:2016kfm},
	\begin{equation}
	E^2 = p^2 c^2 + A_\alpha p^\alpha c^\alpha \,,
	\end{equation}
where $ E $ and $ p $ are the energy and momentum of the GW, and $ A_\alpha $ and $ \alpha $ are phenomenological parameters. The phenomenological parameters are related to the graviton's mass by $m_g = {\sqrt{A_0}}/{c^2}$, with the condition $ A_0 > 0$. 

The 90\% upper bound on the mass of the graviton measured from GW150914 is $ 9.9 \times 10^{-23}  $ eV$ /c^2 $ \cite{LIGOScientific:2019fpa}. If the graviton is indeed massless, then this constraint will get tighter with more detections, especially, with those that turn out to be louder than GW150914.
%should get tighter with detections louder than GW150914, as well as with more detections. 
We assume that the upper bound scales inversely with the SNR of the detection. So, for a given event $ i $ with SNR $\rho_i$, the 90\% upper bound on the graviton mass $ \sigma_{i} $ is given by $	\sigma_{i}  = {\sigma_{\mathrm{0}}} {\rho_\mathrm{0}}/{\rho_{i}}$. Here, $\sigma_{0}$ and $\rho_\mathrm{0} $ are the 90\% upper bound on graviton mass and the SNR obtained from GW150914. If $ N $ events are detected, we can combine the individual $\sigma_{i}$ to obtain the following joint constraint:
	\begin{eqnarray}
    \sigma_\mathrm{comb} = \frac{1}{\sqrt{\sum_{i=1}^{N} \sigma_{i}^{-2} }} = \frac{ \sigma_{0}\,\rho_{0}}{\sqrt{\sum_{i=1}^{N} \rho_i^{2} } } \,.
	\end{eqnarray}

We simulate sources in the HL and AHL networks using the models specified in Sec.~\ref{sec:sim}, and consider only those events that are detected by the multi-detector coincidence criterion (criterion (ii) in Sec.~\ref{sec:rates}). We find that there are, on average, $502$ events per year in the HL network and $754$ events per year in AHL network that satisfy this criterion. We apply the above prescription to obtain the constraints and find ${\sigma_\mathrm{comb}^\mathrm{HLA}}/{\sigma_\mathrm{comb}^\mathrm{HL}} \approx 0.8 $, i.e.\ the constraints obtained with the AHL network are $20\%$ tighter than the ones obtained with the HL network. 
% NUMBERS CAN BE FOUND IN scripts/TGR_improvement_with_HLA.ipynb
\subsection{Constraints on the nature of GW polarisations} \label{sec:pol}
%	\cmt{Assigned to: Srashti Goyal and P. Ajith}

In GR, GWs have only two independent polarisation states --- i.e., two transverse quadrupole (or tensor) modes. In comparison, a general metric theory of gravity can admit up to six polarisation modes. In this sense, GW  polarisations offer an interesting test of GR. GW polarisations can be constrained from observations of spinning neutron stars~\cite{Isi:2017equ,knownPulsars:GWsearch} and stochastic background~\cite{Callister:2017ocg,Abbott:2018utx,Nishizawa:2009bf}, as well as from observations of compact binary mergers~\cite{Isi:2017fbj,Pang:2020pfz,GW17O817:tgr,gw170814,ModelIndependentPolTestKaterina,PurePolInclination,GW170817MixedPol}. While the detectability of spinning neutron stars or stochastic background is uncertain, we are expecting to detect hundreds to thousands of compact binary mergers in the next few years using ground-based GW detectors. Note that each GW detector observes only \textit{one} linear combination of these polarisations. Due to the near co-alignment of the LIGO-Hanford and -Livingston detectors, they measure essentially the \emph{same} linear combination of polarisations in a binary merger signal. Hence, currently the LIGO detectors alone are practically incapable of resolving even the two polarisation states predicted by GR. Additional detectors around the globe, including LIGO-Aundha, will enable observing those two states and potentially constrain the additional non-GR polarisation states.

Given the data from a network of GW detectors, we can compare the posterior probabilities of different hypotheses, for example, one hypothesis stating that the binary phase evolution is exactly as predicted by GR, while the alternative hypothesis accommodating the presence of additional modes~\cite{Isi:2017fbj}. Motivated by the limited number of linearly independent detectors to observe the polarisation modes, the current probes of the nature of GW polarisations have employed highly simplified hypotheses as alternatives to GR. That is, the alternative hypothesis assumes that the polarisations contain only scalar modes ($h_\mathrm{b}$ and $h_\mathrm{l}$) or only vector modes ($h_\mathrm{x}$ and $h_\mathrm{y}$) or only tensor modes ($h_+$ and $h_\times$)~\cite{gw170814,GWTC1TestofGR,GW17O817:tgr,PurePolInclination}. 

We perform a simulation study that compares the ability of the 3-detector network involving LIGO-Aundha to distinguish different polarisation models, as compared to the 2-detector network involving only LIGO Hanford and Livingston.  For each polarisation hypothesis --  tensor $H_t$, vector $H_v$, and scalar $H_s$, the model waveforms are generated using the corresponding antenna patterns, but always assuming that the time evolution of the polarisations follows that of the GR modes. That is, $h_\mathrm{b}(t) = h_\mathrm{x}(t)  = h_+(t) $ and $h_\mathrm{l}(t)  = h_\mathrm{y}(t) = h_\times(t)$. Since each detector $\textit{I}$ has different antenna pattern functions $F^A_I$ for each GW polarisation $A$, the strain measured is a different linear combination of the polarisation modes: $h_I(t) = F_I^A(\alpha, \delta, \psi, t) h_A(t)$. 
%already defined in the PE section: where $\alpha, \delta$ denotes the sky location of the source and $\psi$ the polarisation angle. 
For simplicity, no noise is added to the simulated signals. Further, we considered GW signals from non-spinning binary black holes as our signal model. 

We use the standard Gaussian likelihood model for estimating the posteriors of the parameters under different polarisation hypotheses~(see, e.g.,~\cite{Veitch:2014wba}), using the \textsc{Bilby} software package~\cite{bilby}. Posteriors are computed over the parameters $(m_1, m_2, \alpha, \delta, d_L, \iota, \psi, \phi_0, t_0)$, where $t_0$ and $\phi_0$ are the arrival time and phase, respectively. {We use uniform priors in redshifted component masses of the binary {($m_1, m_2 \in [3,500 ] M_\odot$)}, isotropic priors in sky location (uniform in $\alpha, ~\sin \delta$) and orientation (uniform in $\cos \iota, ~\phi_0$), uniform prior in polarisation angle $\psi$, and a volumetric prior $\propto d_L^2$ on luminosity distance.} The Bayesian evidence of each polarisation model is obtained as part of the parameter estimation. 

\begin{figure}
	\centering
	\includegraphics[width = 0.65 \linewidth]{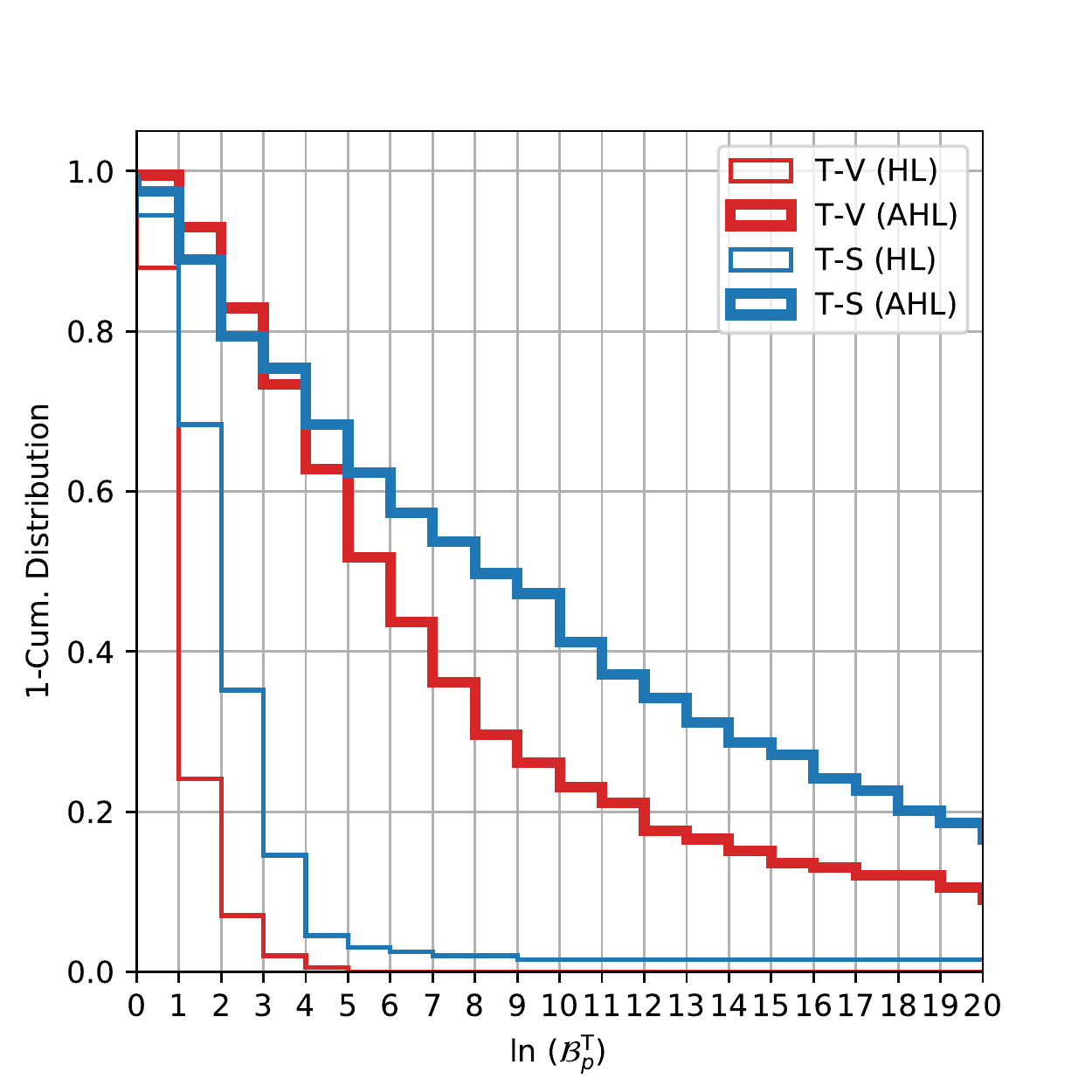}
	\caption{The distribution of the Bayes factors, $B^t_v$(red) and $B^t_s$(blue), that show our ability to distinguish different polarisation models. The simulated signals always follow tensor polarisations (as predicted by GR) and Bayesian evidence of three different polarisation models are computed. We can see that the 3-detector AHL network has a much better ability (larger Bayes factors) to distinguish the polarisation models as compared to the 2-detector HL network(thin).}
	\label{fig-polHLA}		
\end{figure}

We simulate $\sim$200 GR (tensor) double coincident injections for each one of HL and AHL networks, and do parameter estimation and compute evidences $P({d}|H_p)$ for each of the polarisation hypothesis $H_p \in \{H_t,  H_v, H_s\}$. From those evidences, the likelihood ratio (Bayes factor) for tensor {\it vs} vector ($B^t_v$) and tensor {\it vs} scalar ($B^t_s$) hypotheses are calculated for various combinations of detectors. The distributions of the Bayes factors $B^{t}_s := {P({d}|H_t)}/{P({d}|H_s)}$ and $B^{t}_v = {P({d}|H_t)}/{P({d}|H_v)}$ are plotted in Fig.~\ref{fig-polHLA}. We can see that the 3-detector AHL network has a much better ability (larger Bayes factors) to distinguish the polarisation models as compared to the 2-detector HL network.

% \cmt{for the figure, script used: https://git.ligo.org/ligo-india/science_case/-/blob/master/data/POL_data_plot_script/POL-plot.py}

%\begin{figure}
%	\centering
%	\includegraphics[width = 0.65 \linewidth]{POL-network.pdf}
%	\caption{Bayesfactors distribution for various detector networks.}
%	\label{fig-polnetwork}		
%\end{figure}

%\input{stoch}
%\input{polarisation}

\section{Conclusions and summary}
\label{sec:conclusion}

%{\color{blue} Writers of PE, Rates, TGR sections should add short summaries below.}

We have analyzed the performance of the LIGO Global Network, in particular focusing on the improvement that comes about from the addition of a LIGO detector in India. We focused on compact binary coalescences, involving black holes and neutron stars, as our sources. We find, overall, a significant improvement in the precision with which various binary parameters can be measured. This is especially significant for the sky localization of compact binaries as well as the related ability to issue early warning for BNS mergers. Precise localization is crucial in spotting and associating kilonovae with such mergers. Correct associations are necessary for understanding the influence of progenitor properties on kilonova properties, such as their spectra. They also have a bearing on constraining neutron star equation of state and the measurement of the Hubble parameter without invoking the cosmic-distance ladder.

Reducing the alert time for these mergers by several to a few tens of seconds can impact the ability of astronomers to slew their telescopes in time to capture prompt afterglow emissions, not to mention pre-merger EM signals. Prompt afterglows were missed in the observation of GW170817 and can provide important clues about the short GRB engine. This is one of the next frontiers in GRB physics that the GW network can contribute to.

The increased detection rate of CBCs with the addition of the third LIGO detector in India will also allow stronger constraints to be placed on possible deviations from GR. By assuming that constraints from a given signal would scale inversely with the SNR, and combining constraints across events, we find that the AHL network will offer a $20\%$ improvement over the HL network on the graviton mass upper limit. We also showed how an additional detector in the network 
aids in discriminating among different polarisation models:
%enable the measument of additional linear combinations of polarisations --- we find that the Bayes %Factor in AHL will be four time stronger as opposed to HL. 
Here we limited ourselves to models where GWs have only tensor, only vector or only scalar polarisation modes -- taking all modes to have the same phase evolution; some of these assumptions can be relaxed in the future by using the null stream reconstruction~\cite{ModelIndependentPolTest, ModelIndependentPolTestKaterina,GW170817MixedPol}.

% and the improvement in cosmological inference are particularly impressive and will allow for a new picture of the universe.

While our primary focus here has been CBCs, the LIGO Global Network will also impact science pursuable with other signals. One such signal is an astrophysical stochastic gravitational-wave background (SGWB) arising from the superposition of inspiral signals from populations of binaries of black holes and neutron stars~\cite{ASGWB_Regimbau_2008,ASGWB_Regimbau_2011}. By combining the detections of dozens of individual binaries, on the one hand, with upper-limits on the power spectra of SGWB, on the other hand, past studies~\cite{LVK_O3Isotropic,CallisterSGWB} have constrained the rate of evolution of CBCs over redshift. The spread in this rate will shrink by 20\% with the third detector added to this network. The addition of a third detector will also help to understand the correlated terrestrial noise sources, which in turn will play a crucial role in confidently claiming any SGWB detection. Moreover, if the astrophysical SGWB has significant anisotropies, probing them requires better sky coverage. LIGO-India's inclusion in the existing detector network will help to resolve these finer angular structures using the existing mapping techniques~\cite{GWRadiometer_Mitra, MultiBaseline_Talukder,pystoch_2018,sph_pystoch_2020}.

In the future, it will be interesting to study other types of sources (and not just compact binaries), the effects of realistic interferometer noise, and the presence of other detectors in the network.

\section*{Acknowledgements}
We would like to thank our colleagues in the LIGO-India Scientific Collaboration and the LIGO-India Project for valuable inputs. We appreciate the several discussions we had with members of the various working groups in the LIGO-Virgo-KAGRA collaborations. In particular, we thank K.~G.~Arun, Bala Iyer, Shivaraj Kandhasamy, Jose Matthew, Fred Raab, Rory Smith, and Tarun Souradeep for valuable discussions and inputs.  This work makes use of \textsc{NumPy} \cite{vanderWalt:2011bqk}, \textsc{SciPy} \cite{Virtanen:2019joe}, \textsc{Matplotlib} \cite{Hunter:2007}, \textsc{AstroPy} \cite{Robitaille:2013mpa,Price-Whelan:2018hus}, \textsc{jupyter} \cite{jupyter}, \textsc{dynesty} \cite{2019S&C....29..891H}, \textsc{bilby} \cite{bilby} and \textsc{PESummary} \cite{Hoy:2020vys} software packages. Thanks are also due to the Department of Science and Technology (DST) and the Department of Atomic Energy (DAE) of India. Specifically, MS acknowledges the support from the Infosys Foundation, the Swarnajayanti fellowship grant DST/SJF/PSA-01/2017-18, and the support from the National Science Foundation with grants PHY-1806630 and PHY-2010970, AP acknowledges support from the DST-SERB Matrics grant MTR/2019/001096 and SERB-Power-fellowship grant SPF/2021/000036, DST,India. AM acknowledges support from the DST-SERB Start-up Research Grant SRG/2020/001290, and PA, AV, and SG acknowledge support from DAE under project no. RTI4001.
PA's research was also supported by the Max Planck Society through a Max Planck Partner Group at ICTS-TIFR and by the Canadian Institute for Advanced Research through the CIFAR Azrieli Global Scholars program. SS is supported by an Eberly postdoctoral fellowship at Pennsylvania State University and B.S.S. is supported by NSF grants PHYS-1836779, PHYS-2012083 and AST-2006384. Thanks are due to computational support provided by the Alice (ICTS-TIFR) and Sarathi (IUCAA) clusters and computing resources in SINP. In addition, the authors are also grateful for the computational resources provided by LIGO Laboratory and Leonard E Parker Center for Gravitation, Cosmology and Astrophysics at the University of Wisconsin-Milwaukee and supported by National Science Foundation Grants PHY-0757058, PHY-0823459, PHY-1626190 and PHY-1700765.
This paper has been assigned the internal LIGO preprint number P2100073.

\clearpage
\bibliographystyle{iopart-num}
\bibliography{references}
\end{document}